%% file: Main.tex
\documentclass[twocolumn]{aastex63}

\usepackage{array}
\newcolumntype{P}[1]{>{\centCas Aering\arraybackslash}p{#1}}
\usepackage{supertabular}
\usepackage{graphicx}
\usepackage{cellspace}
\usepackage{amssymb}
\usepackage{amsmath}
\usepackage{longtable}
\usepackage{tabu}
\usepackage{color}
\usepackage{etoolbox}
\usepackage{supertabular}
\usepackage{dcolumn}
\usepackage{ wasysym }

\input{macros}

\newcommand\Tstrut{\rule{0pt}{2.9ex}}       
\newcommand\Bstrut{\rule[-1.3ex]{0pt}{0pt}} 
\newcommand\TBstrut{\Tstrut\Bstrut}  

\newtoggle{fullauthorlist}
\toggletrue{fullauthorlist}

\newtoggle{endauthorlist}
\toggletrue{endauthorlist}

\shorttitle{ 
Einstein@Home search for Continuous Gravitational Waves from Vela Jr. and G347.3
}
\shortauthors{Ming et al.}
\begin{document}

\title{ 
Deep Einstein@Home search for Continuous Gravitational Waves from the Central Compact Objects in the Supernova Remnants Vela Jr. and G347.3-0.5 using LIGO public data
}

\author{J. Ming}
\email{jing.ming@aei.mpg.de}
\affiliation{Max Planck Institute for Gravitational Physics (Albert Einstein Institute), Callinstrasse 38, D-30167 Hannover, Germany}
\affiliation{Leibniz Universit\"at Hannover, D-30167 Hannover, Germany}

\author{M.A. Papa}
\email{maria.alessandra.papa@aei.mpg.de}
\affiliation{Max Planck Institute for Gravitational Physics (Albert Einstein Institute), Callinstrasse 38, D-30167 Hannover, Germany}
\affiliation{Leibniz Universit\"at Hannover, D-30167 Hannover, Germany}

\author{H.-B. Eggenstein}
\affiliation{Max Planck Institute for Gravitational Physics (Albert Einstein Institute), Callinstrasse 38, D-30167 Hannover, Germany}
\affiliation{Leibniz Universit\"at Hannover, D-30167 Hannover, Germany}

\author{B. Beheshtipour}
\affiliation{Max Planck Institute for Gravitational Physics (Albert Einstein Institute), Callinstrasse 38, D-30167 Hannover, Germany}
\affiliation{Leibniz Universit\"at Hannover, D-30167 Hannover, Germany}

\author{B. Machenschalk}
\affiliation{Max Planck Institute for Gravitational Physics (Albert Einstein Institute), Callinstrasse 38, D-30167 Hannover, Germany}
\affiliation{Leibniz Universit\"at Hannover, D-30167 Hannover, Germany}

\author{R. Prix}
\affiliation{Max Planck Institute for Gravitational Physics (Albert Einstein Institute), Callinstrasse 38, D-30167 Hannover, Germany}
\affiliation{Leibniz Universit\"at Hannover, D-30167 Hannover, Germany}

\author{B. Allen}
\affiliation{Max Planck Institute for Gravitational Physics (Albert Einstein Institute), Callinstrasse 38, D-30167 Hannover, Germany}
\affiliation{Leibniz Universit\"at Hannover, D-30167 Hannover, Germany}
\affiliation{University of Wisconsin Milwaukee, 3135 N Maryland Ave, Milwaukee, WI 53211, USA}

\author{M. Bensch}
\affiliation{Max Planck Institute for Gravitational Physics (Albert Einstein Institute), Callinstrasse 38, D-30167 Hannover, Germany}
\affiliation{Leibniz Universit\"at Hannover, D-30167 Hannover, Germany}

\begin{abstract}

We perform a search for continuous nearly monochromatic gravitational waves from the central compact objects associated with the supernova remnants  Vela Jr. and  G347.3 using LIGO O2 and O3 public data. Over $10^{18}$  different waveforms are considered, covering signal frequencies between 20-1300 Hz (20-400 Hz) for G347.3-0.5 (Vela Jr) and a very broad range of frequency derivatives. 
Thousands of volunteers donating compute cycles through the computing project Einstein@Home have made this endeavour possible.
Following the Einstein@Home search, we perform multi-stage follow-ups of over 5 million waveforms. 
The selection threshold is set so that a signal could be confirmed using the first half of the LIGO O3 data. 
We find no significant signal candidate for either targets. 
Based on this null result, for G347.3-0.5, we set the most constraining upper limits to date on the amplitude of gravitational wave signals, corresponding to deformations below $10^{-6}$ in a large part of the search band.  At the frequency of best strain sensitivity, near $161$ Hz, we set 90\% confidence upper limits on the gravitational wave intrinsic amplitude of $h_0^{90\%}\approx 6.2\times10^{-26}$. Over most of the frequency range our upper limits are a factor of 10 smaller than the indirect age-based upper limit. For Vela Jr.,  near $163$ Hz, we set $h_0^{90\%}\approx 6.4\times10^{-26}$. Over most of the frequency range our upper limits are a factor of 15 smaller than the indirect age-based upper limit. The Vela Jr. upper limits  presented here are slightly less constraining than the most recent upper limits of \cite{ligo_o3a_c_v} but they apply to a broader set of signals. 
\end{abstract}


\section{Introduction}
\label{sec:introduction}

Continuous gravitational waves are nearly monochromatic and long-lasting signals. Despite the simplicity of the waveforms, the search for continuous wave signals is highly challenging task due to the extreme weakness of waves' amplitude. Integrating the signal over many months is necessary to accumulate a sufficient signal-to-noise ratio. However, this also means that when the signal waveform is not known in advance, a large number of different waveforms must be searched for, resulting in a significant increase in computing costs. In fact, the sensitivity of continuous wave searches is typically limited by computing power when searching for a broad range of waveforms. Continuous wave signals remain undetected yet.

Various physical mechanisms could give rise to continuous gravitational waves:
fast-spinning neutron stars (NSs) with non-axisymmetric deformations or unstable r-modes \citep{Owen:1998xg,Owen:2010ng,lasky_2015}, fast inspiral of dark-matter objects  \citep{Horowitz:2019pru, Horowitz:2019aim} and superradiant emission of axion-like particles around black holes \citep{Arvanitaki+2015,Zhu_2020}.  Among these  candidates,  fast spinning  neutron stars are generally believed to be the most promising continuous wave sources.

Many searches for continuous gravitational wave signals have been carried out on advanced LIGO  \citep{2015aligo}:  
The most sensitive and computationally inexpensive searches are those focused on pulsars with known spin frequency and frequency evolution \citep{LVC_2019_targeted,LVC_2020_targeted,Ashok:2021fnj, LVC_2022_targeted}.  On the other hand, all-sky searches are the most computationally expensive due to the lack of any prior information about frequency and sky location  \citep{Dergachev:2020fli, Dergachev:2021ujz, Dergachev:2020upb,Steltner_2021, lvc_o2_as, lvc_o3_as_binary, LIGOScientific:2021tsm, O2AS_binary, ligo_o3_as, dergachev2022frequency, Covas_2022,steltner2023deep, Singh:2022hfd, LIGOScientific:2021yby}.

Directed searches lie in-between these two extremes, with known sky positions, but unknown spin frequency. This type of searches include  interesting spots such as the galactic centre or the globular cluster Terzan 5 \citep{Piccinni_o2gc,Dergachev:2019pgs,ligo_o3gc},  thought to host a large number of neutron stars, young supernova remnants  \citep{Ming:2019xse, Papa_2020midth, lvc_o1_snr, Abbott:2018qee,Millhouse_o2_snr,Lindblom:2020rug,LIGOScientific:2021mwx, ligo_o3gc, ligo_o3a_c_v, Owen_2022, liu_2022, Ming_2022, O3_deep_C},  and low-mass X-ray binaries such as Scorpius X-1 \citep{Zhang_2021, Whelan_2023}.

The young central compact objects found in supernova remnants are particularly interesting for several reasons. 
The young neutron stars spin down rapidly enough to potentially emit relatively strong continuous waves. 
Besides,  theories suggest that r-mode oscillations in neutron stars could be detectable in stars that are up to several thousand years old \citep{Owen:2010ng}.
For these reasons among 294 Galactic supernova remnants \citep{Green_2019},  15 young supernova remnants have been considered as emitters of continuous gravitational waves \citep{Abbott:2018qee, LIGOScientific:2021mwx} and searches have been carried out to detect gravitational waves from them. 

Since no rotation frequency measurements exist for any of these objects, the gravitational signals could appear at virtually any frequency that the ground-based detectors can ``see", i.e. between 20-2000 Hz. The frequency evolution could also be quite diverse, depending on the braking mechanisms at play. This results in a very large number of waveforms to probe and hence to a significant computational cost, at least for high-sensitivity searches. 

The question then arises of how to pick among the 15 young supernova remnants, and what frequency and frequency-derivative range to cover, and with what search, in order to maximise the detection probability at fixed computational resources.  In \citep{Ming:2015jla} we propose an optimisation scheme to address this question. The main practical result is that, among the 15 young supernova remnants, Vela Jr. (G266.2-1.2), Cassiopeia A (G111.7-2.1), and G347.3 (G347.3-0.5) are most worthy to invest in, with extensive searches.

Guided by this study, in the quest to detect a signal, a number of searches have been performed targeting these objects on more and more sensitive data. We continue in this vein by extending to higher frequencies and to higher sensitivity previous searches for emission from G347.3 \citep{Ming:2019xse,Papa_2020midth,Ming_2022}. We also search for emission from Vela Jr, with a different method, a broader frequency-evolution model, and a different data set than the latest search for emission from this object, reported by \cite{ligo_o3a_c_v}. Despite of us using data about a factor 1.6 less sensitive, our sensitivity to Vela Jr. is less than 20\% worse than that of \cite{ligo_o3a_c_v}.

This paper is organised into following sections: in Section \ref{sec:target} we give a brief review of the astrophysical targets and the gravitational waveform model. 
In section \ref{sec:Data}, we describe the data used in this work; in section \ref{sec:search} the semi-coherent search methodology; in section \ref{sec:multi_stage}, we detail the follow-up searches; in Section \ref{sec:results}, we present the results which we discuss in Section \ref{sec:conclusions}.

\section{The target and signal}
\label{sec:target}

\subsection{G347.3-0.5}
\label{sec:G347.3}

Among the supernova remnants in our galaxy, G347.3 is one of the most interesting targets for continuous gravitational waves because of its relatively young age and close distance \citep{Ming:2015jla}. 

The supernova remnant G347.3 is suggested to be the remnant of the AD393 ``guest star" \citep{1997A&A...318L..59W}. We therefore assume an age of 1600 years, albeit this estimate is not completely uncontroversial \citep{2012AJ....143...27F}. Using XMM data, \cite{2004A&A...427..199C} estimate its distance to be around 1.3 kpc. The position of the central compact object in the G347.3 supernova remnant is given with sub-arcsecond accuracy by \cite{2008A&A...484..457M}, based on Chandra data.

\subsection{Vela Jr.}
\label{sec:velajr}

Similar to G347.3, Vela Jr. is a promising target because of its relatively young age and close distance, especially in one of the scenarios described below \citep{Ming:2015jla}.

The supernova remnant Vela Jr. is also known as G266.2-1.2. It is located within the Vela constellation in the southern part of the sky.
It is referred to as Vela Jr because it is enclosed by the much larger and older supernova remnant Vela \citep{1968Natur.220..340L}.
The sky position with sub-arcsecond accuracy corresponds to compact central object discovered with Chandra data \citep{Pavlov_2001}. Both age and distance of Vela Jr. are uncertain and  two extreme scenarios are considered in this work. We utilize the smaller age of 700 years and smaller distance of 200 pc as estimated by \cite{iyudin_1998}, corresponding to the most favourable a-apriori conditions for the detectability of a continuous signal.
For the least favourable scenario we use an age  of 5100 years from \citep{Allen_2014}.  Those authors also explore the potential association of several neighbouring objects with the closer concentration of the Vela Molecular Ridge and obtain an estimated distance of $\le1$ kpc  which is consistent with the previous work of \cite{Liseau1992}, who estimate a distance of $0.7\pm0.2$ kpc.  In this work, we take $0.9$ kpc to be the pessimistic distance.

\subsection{The Signal}
\label{sec:signal}

In the detector data, the continuous wave signal produced by an asymmetric rotating neutron stars takes the form \citep{Jaranowski:1998qm}:
\begin{equation}
h(t)=F_+(t)h_+(t)+F_{\times}(t)h_{\times}(t),
\label{eq:signal}
\end{equation}
where $F_+(t)$ and $F_\times(t)$ are  the detector beam-pattern functions for the two gravitational wave polarizations ``+" and ``$\times$".
They depend on the sky position of the source (defined by the right ascension $\alpha$ and
declination $\delta$), and the orientation $\psi$ of the wave-frame with respect to the detector frame.
 $F_+(t)$ and $F_\times(t)$ are periodic time functions with a period of one sidereal day, because the detector rotates with the Earth.
The waveforms for the two polarizations, $h_+(t)$ and $h_\times(t)$ can be expressed as: 
\begin{eqnarray}
h_+ (t)  =  A_+ \cos \Phi(t) \nonumber \\
h_\times (t)  =  A_\times \sin \Phi(t).
\label{eq:monochromatic}
\end{eqnarray}
Here,  $\Phi(t)$ is  the phase of the gravitational-wave signal at the time $t$ and  $A_{+,\times}$ are the polarizations' amplitudes which take the form:
\begin{eqnarray}
A_+  & = & {1\over 2} h_0 (1+\cos^2\iota) \nonumber \\
A_\times & = &  h_0  \cos\iota. 
\label{eq:amplitudes}
\end{eqnarray}
Here $h_0$ is the intrinsic gravitational wave amplitude and $\iota$ is the angle between the total angular momentum of the star and the line of sight.

The phase $\Phi(t)$ of the signal at the solar system barycenter (SSB) frame has the form:
\begin{multline}
\label{eq:phiSSB}
\Phi(\tau_{\mathrm{SSB}}) = \Phi_0 + 2\pi [ f (\tau_{\mathrm{SSB}}-{\tau_0}_{\mathrm{SSB}})  +
\\ {1\over 2} \dot{f} (\tau_{\mathrm{SSB}}-{\tau_0}_{\mathrm{SSB}})^2 + {1\over 6} \ddot{f} (\tau_{\mathrm{SSB}}-{\tau_0}_{\mathrm{SSB}})^3 ], 
\end{multline}
where $f$ is the signal frequency and $\tau_{\mathrm{SSB}}$ is the arrival time of the GW front at the SSB frame and ${\tau_0}_{\mathrm{SSB}}$ is a reference time.

\section{The data}
\label{sec:Data} 

We only use data from the LIGO detectors, because the sensitivity of the Virgo detector is not sufficient to offset the additional computational cost of analysing a third data stream.  In other words, within a constrained computational budget the most sensitive search that one can perform entails only the two LIGO detectors. 

We use public data from the second and the third observing runs, O2 and O3a \citep{Abbott:2021boh, Abbott_2023}.
The O2 data span is between GPS time 1167983370 (Jan 09 2017) and 1187731774 (Aug 25 2017) and the O3a data span between GPS time 1238421231 (Apr 04 2019) and 1253973231 (Oct 01 2019). The reference time we adopt in the search using O2 data $\tau_{\textrm{SSB}}^{\textrm{O2}}$ is 1177858472.0, and for the O3a data,  $\tau_{\textrm{SSB}}^{\textrm{O3a}}$ is 1246197626.5.

While the whole O3 data set is publicly available, we use only the first part of O3 (O3a). The reason is that O3a is enough to weed-out candidates due to noise fluctuations, but a signal would require an independent data set for confirmation, and the second half of the O3 data provides that. This impacts the sensitivity slightly, but, with no public data access to more data, such as the O4 data, until 2025, it is the best that can be done.

Short Fourier transforms (SFTs) of data segments 1800 seconds long  \citep{SFTs} are created as customary for \EatH \, searches and are used as input to the search. Calibration lines, the mains power lines and some other spurious noise due to the LIGO laser beam jitter are removed in the publicly released O2 and O3 data  \citep{Davis_2019,Vajente_2020}. Additionally we remove loud short-duration glitches with the gating procedure described in  \cite{Steltner:2021qjy} and substitute Gaussian noise in the frequency bins affected by lines \citep{line_list_o2}.
The line files we adopt for O2 data can be found at \citep{o2_linefile}.  For O3a data set, the line file is the file in version 1.7 which can be found at \citep{o3_linefile}.

\section{The Searches}
\label{sec:search}

\subsection{Semi-coherent searches}

The main building block in this work is a ``stack-slide"  type of search based on the GCT (Global correlation transform) method  \citep{PletschAllen,Pletsch:2008,Pletsch:2010}. 
The whole data with a span $\Tobs$ is split in $\Nseg$ segments and each segment spans a duration $\Tcoh$. The data of both detectors from each segment $i$ is searched with a maximum likelihood coherent method to construct the detection statistic, $\F$-statistic  \citep{Cutler:2005hc}. 
The results from these coherent searches are combined by summing these  $\F_i$-statistic values from different segments and this yields
the value of the core detection statistic $\avF$: 
\begin{equation}
\label{eq:avF}
\avF:={1\over\Nseg} \sum_{i=1}^{\Nseg} \F_i.
\end{equation}
 
In Gaussian noise $\Nseg\times 2\avF$ follows a chi-squared distribution with $4\Nseg$ degrees of freedom. 
If a  signal is present, this chi-squared distribution has a non-centrality parameter 
\begin{equation}
\label{eq:rho2}
\rho^2\propto {h_0^2\Tobs \over {S_h}},
\end{equation}
where $S_h$ is the strain power spectral density of the noise at the frequency of the signal \citep{Jaranowski:1998qm}.

The $\F_i$-statistic is a coherent combination of the data from all detectors in the $i$-th segment, whereas the summation of the detection statistic values in Eq.~\ref{eq:avF} is not. This is the reason why this type of search is called ``semi-coherent".

Despite removing loud glitches and lines, some coherent disturbances may persist. These are not astrophysical signals but they are more signal-like than Gaussian noise, and hence they give rise to increased values  of $\avF$. To decrease these false alarms a ``line-robust" detection statistic $\BSNtsc$ is computed using the log of a Bayesian odds ratio that compares the signal hypothesis to an extended noise hypothesis  \citep{Keitel:2013wga,Keitel:2015ova}. This noise model accounts for not only Gaussian noise but also for coherent disturbances. 

To increase the computational efficiency of the semi-coherent search, the initial detection statistic value is an approximation of the exact value for any given template. If the candidate is due to a signal, the non-approximated detection statistic, exactly computed for the template, will in general yield a higher detection statistic value than the approximated one. For this reason, after the search is done, the detection statistic of the highest-ranking results is recomputed at the exact template. The resulting values are indicated with a subscript ``r", indicating that they are quantities {\it{r}}ecalculated at the exact template: $2\F_r$  and  $\BSGLtLr$.

We take a hierarchical approach, and carry out a series of semi-coherent searches. The sensitivity of the searches increases as the stages progress, so that the significance of a signal increases as it is shows up in the results of more and more advanced stages. Conversely the significance of noise fluctuations or disturbances in general does not increase as for a signal, and we use this fact to ``weed-out" spurious candidates.  So the number of candidates that survive from one stage to the next decreases, and at the end we are left with only the most significant ones, if any.

Important variables for a semi-coherent search are: the coherent time baseline of the segments $\Tcoh$, the number of segments used $\Nseg$, the total time $\Tobs$ spanned by the data, the grids in parameter space and the detection statistic used to rank the parameter space cells. These parameters are given in Table \ref{tab:GridSpacings}.  

The first stage is the most computationally expensive because the entire parameter space is searched. Based on the detection statistic values, the most promising candidates are identified and only those are passed on to the second stage. In the second stage an appropriate volume of parameter space is searched around the nominal candidate parameters, and if the outcome is consistent with what expected from a signal, the candidate passes on to the third stage, else it is discarded. In total we have four stages, numbered from 0 to 3.

\subsection{Stage 0}
\label{sec:O2search}

Stage 0 is the $\mathrm{\EatH}$ search.  $\mathrm{\EatH}$ is built on the BOINC (Berkeley Open Infrastructure for Network Computing) architecture~ \citep{Boinc2,Boinc3} where  volunteers use their computers in idle time  to tackle scientific problems such as this, that require large amounts of computing power.

We search for signal-waveforms with frequency and frequency-derivatives as follows:
\begin{equation}
\label{eq:Priors}
	\begin{cases}
	20 ~\mathrm{Hz} \le f  \le 1300 ~\mathrm{Hz}~~~~~~~\textrm{G347.3} \\
	20 ~\mathrm{Hz} \le f  \le 400 ~\mathrm{Hz}~~~~~~~~\textrm{Vela Jr.} \\
	-f/ \tau\,  \le   \dot{f}  \le 0\,~\mathrm{Hz/s}\\
	0\,\mathrm{Hz/s}^2 \leq  \ddot{f} \leq ~7\dot{|f|}_{\textrm{max}}^2/f = 7 {f/\tau^2}.
	\end{cases}
\end{equation}
The ranges for $\dot{f}$ and $\ddot{f}$ of Eq.s~\ref{eq:Priors} correspond to different breaking index $n$ values, namely 2 and 7:
In the $\dot{f}$-range equation the lower bound is $-\frac{f}{\tau}\frac{1}{n-1}$ so the lowest value corresponds to a frequency evolution with a braking index $n=2$, which is what we have assumed in order to encompass the broadest range of $\dot{f}$ values. In the $\ddot{f}$  equation the upper bound is proportional to $n\frac{f}{\tau^2}$, so we have assumed $n=7$ to maximise the $\ddot{f}$ range. We have adopted  $\tau=1600$ years for G347.3 and $\tau=700$ years for Vela Jr, that again maximises the search ranges. The search ranges for two targets at 400 Hz are given in Table \ref{tab:SearchParams}.

The duration of the coherent baseline $\Tcoh$, the template grid spacings and the search ranges are all derived from the optimisation procedure \citep{Ming:2015jla}, and are shown shown in Tables \ref{tab:SearchParams} and \ref{tab:GridSpacings}. The template grid spacings are defined by $\delta{f}$, $\delta{\dot{f}}$, $\delta{\ddot{f}}$,  $\gamma_1$ and $\gamma_2$.  The $\delta{f}$, $\delta{\dot{f}}$, $\delta{\ddot{f}}$ are the coarse grid spacings and they are the grid spacing used in the coherent search in each segment. In the incoherent summing part, the coarse grids are refined by a factor of $\gamma_1$ for ${\dot{f}}$ and $\gamma_2$ for ${\ddot{f}}$. 

\begin{figure}[h!tbp]
  \includegraphics[width=\columnwidth]{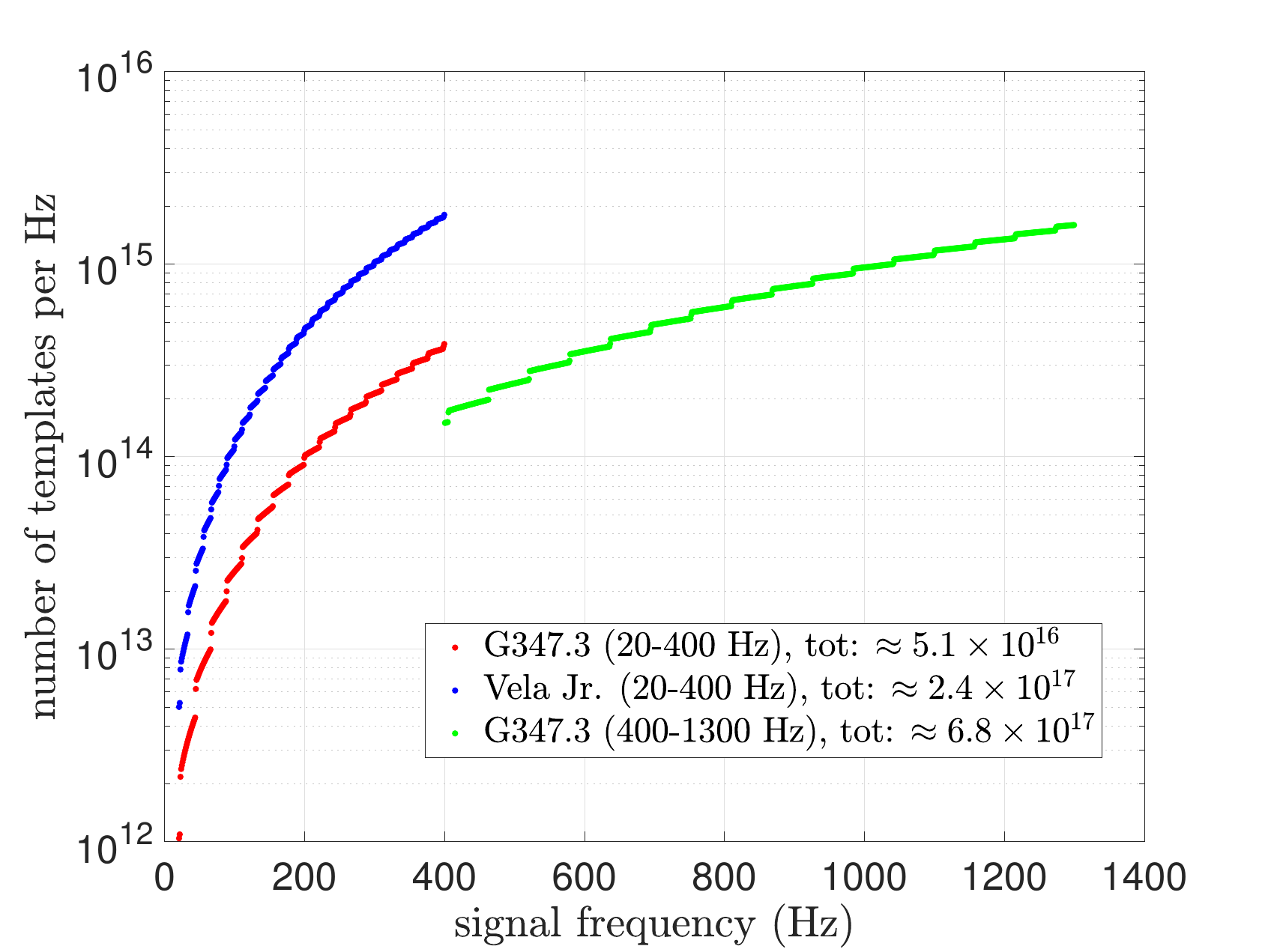}
\caption{Number of templates searched in 1-Hz bands as a function of signal frequency for different searches. In the legend we also show the total number of templates searched for each search.}
\label{fig:HowManyTemplates}
\end{figure}

The number of templates searched at a given frequency varies as a function of frequency and is different for the two targets, due to the different ages (Eq.~\ref{eq:Priors}).
Fig. \ref{fig:HowManyTemplates} shows the number of templates searched in 1-Hz bands as a function of frequency for the two targets.

\begin{table*}[ht]
\centering
\begin{tabular}{|c|c|c|}
\hline
\hline
 & Vela Jr. & G347.3 \\
\hline
\hline
 \TBstrut$f$ range &  {[20-400] Hz} & {[20-1300] Hz} \\
 \hline
  \TBstrut$\Tref$  & \multicolumn{2}{c|} {\TrefGPS GPS s } \\
  \hline
\TBstrut$\fdot $ range  & [\paramfdotloVela - \paramfdothiVela~ ]  Hz/s & [\paramfdotloG - \paramfdothiG ~] Hz/s \\
 \hline
 \TBstrut$\fddot $ range  & [ \paramfddotloVela~ - \paramfddothiVela ~] Hz/s$^2$   & [ \paramfddotloG~ - \paramfddothiG~ ] Hz/s$^2$  \\
  \hline
\TBstrut$ \alpha$ & \posVelaJra & \posGa \\  
\TBstrut$ \delta$  & \posVelaJrd & \posGd \\
 \hline
\hline
\end{tabular}
\caption{Search ranges. The spindown ranges quoted are the ones used at 400 Hz. The ranges at different frequencies are readily derived from Eq.~\ref{eq:Priors}. }
\label{tab:SearchParams}
\begin{tabular}{|c|c|c|c|}
\hline
\hline
 & Vela Jr. (20 - 400 Hz) & G347.3 (20 - 400 Hz) & G347.3 (400 - 1300 Hz) \\
\hline
\hline
 \TBstrut$\Tcoh$ & 720 hr & 1080 hr& 720 hr\\
 \hline

\TBstrut$\Nseg$ & 6 & 5 & 6 \\
  \hline
\TBstrut$\delta f$ & $1.90 \times 10^{-7}$ Hz & $1.27 \times 10^{-7}$ Hz & $1.90 \times 10^{-7}$ Hz \\
 \hline
 \TBstrut$\delta {\dot{f_c}}$ &  $4.49\times 10^{-13}$ Hz/s  & $2.00\times 10^{-13}$ Hz/s  & $4.49 \times 10^{-13}$ Hz/s \\
  \hline
\TBstrut$\gamma_1$ & 21 &  13 & 21 \\
  \hline
\TBstrut$\delta {\ddot{f_c}}$ & $ 1.59\times 10^{-19} ~{\textrm{Hz/s}}^2 $ & $6.08\times 10^{-20}~{\textrm{Hz/s}}^2 $ & $1.59 \times 10^{-19}~ {\textrm{Hz/s}}^2$ \\
  \hline
\TBstrut$\gamma_2$ & 11 & 5 & 11 \\
 \hline
\TBstrut$\avMis$ & 13.7\% & 13.3\%  & 13.7\%  \\

 \hline
\hline  

\end{tabular}
\caption{Spacings on the signal parameters used for the templates in the \EatH~ search (Stage 0) and the average mismatches $\avMis$.}
\label{tab:GridSpacings}

\end{table*}

Overall we search $\approx {10^{18}} $ templates, utilizing \EatH \, for several months. The work-load is split into work-units, sized to keep the average volunteer host busy for {\paramWUcputimeHours} hours on CPUs and {\paramWUgputimemins} minutes on GPUs. The search is split into over {\paramtotalWUsmillionsGV} million work-units and each of them searches $\approx$ {\avgTempWU} templates. The most promising 10\,000 results from each work-unit constitute the so-called ``top-lists" and are communicated back to the central $\mathrm{\EatH}$ server for further post-processing. The total number of results returned from this search is $3\times 10^{10}$, which is a very small fraction of the total number of searched waveforms.

\section{Hierarchical follow up searches}
\label{sec:multi_stage}

Our hierarchical search comprises 4 stages. Stage 0 described above is the $\mathrm{\EatH}$ search. 
At each later stage, we run a search around candidates with a coherent length that increases until Stage 2, where the search is fully coherent. 
At each stage, the top ranking candidates are marked and then searched in the next stage. If the data harbours a real signal, the significance of the recovered candidate will increase with respect to the significance that it had in the Stage 0. On the other hand, if the candidate is not produced by a continuous wave signal, the significance is not expected to increase consistently over the successive stages.

Since Stage 0 already utilises long coherence time-baselines, already yielding only 5 and 6 segments respectively across the different searches, for Stage 1 we simply take intermediate baselines yielding 2 segments. We don't use a fully coherent search because of the computing cost. 
Stage 1 is necessary as it not only reduces the number of candidates that need to be followed up in Stage 2, but it also decreases the uncertainty surrounding the parameters of these candidates, thereby shrinking the search volume for Stage 2. An interesting discussion of the choice of coherence-ladder can be found in \citep{mirasola2024}, in the context of non-deterministic follow-ups. Our choices are broadly consistent with \cite{mirasola2024}'s criteria.
Since the fully coherent search is the most sensitive search we can do using O2 data, the remaining candidates after Stage 2 are investigated with the O3a dataset. This is Stage 3. 

For each follow-up stages, the grid spacings are determined to minimize the computational cost at fixed mismatch. The mismatch value is chosen so that the follow-up can be performed in a reasonable amount of time.
The coherent time baselines and grid spacing details for Stages 1-3 are shown in  Table \ref{tab:GridSpacingsall}.

The follow-up search is performed in a small box in parameter space around each candidate. The highest detection statistic result $2\F_r$ from the results in each box is stored and taken as the new representative for that candidate for that stage. 

Some candidates are discarded and not considered for further follow-up:
\begin{equation}
\textrm{if}~ R^{\textrm{(a)}}< R^{\textrm{(a)}}_{\textrm{thr}}\longrightarrow{\textrm{candidate discarded}},
\end{equation}
where $a$ is the stage index, and 
\begin{equation}
R^{\textrm{(a)}}:={{2\avF^{\textrm{(a)}} - 4 } \over {{2\avF^{\textrm{(0)}} - 4}}}.
\label{eq:Rdef}
\end{equation}
The superscript ``0'' indicates that the detection statistic value comes from the original $\mathrm{\EatH}$ search. For signals the expected value of $R^{\textrm{(a)}}\propto \Tcoh^{(a)}/\Tcoh^{(0)}$ (see Eq.~\ref{eq:rho2}), so we expect it to  increase as a signal candidate is investigated with longer and longer $\Tcoh$. 

\begin{figure}[h!tbp]
   \includegraphics[width=\columnwidth]{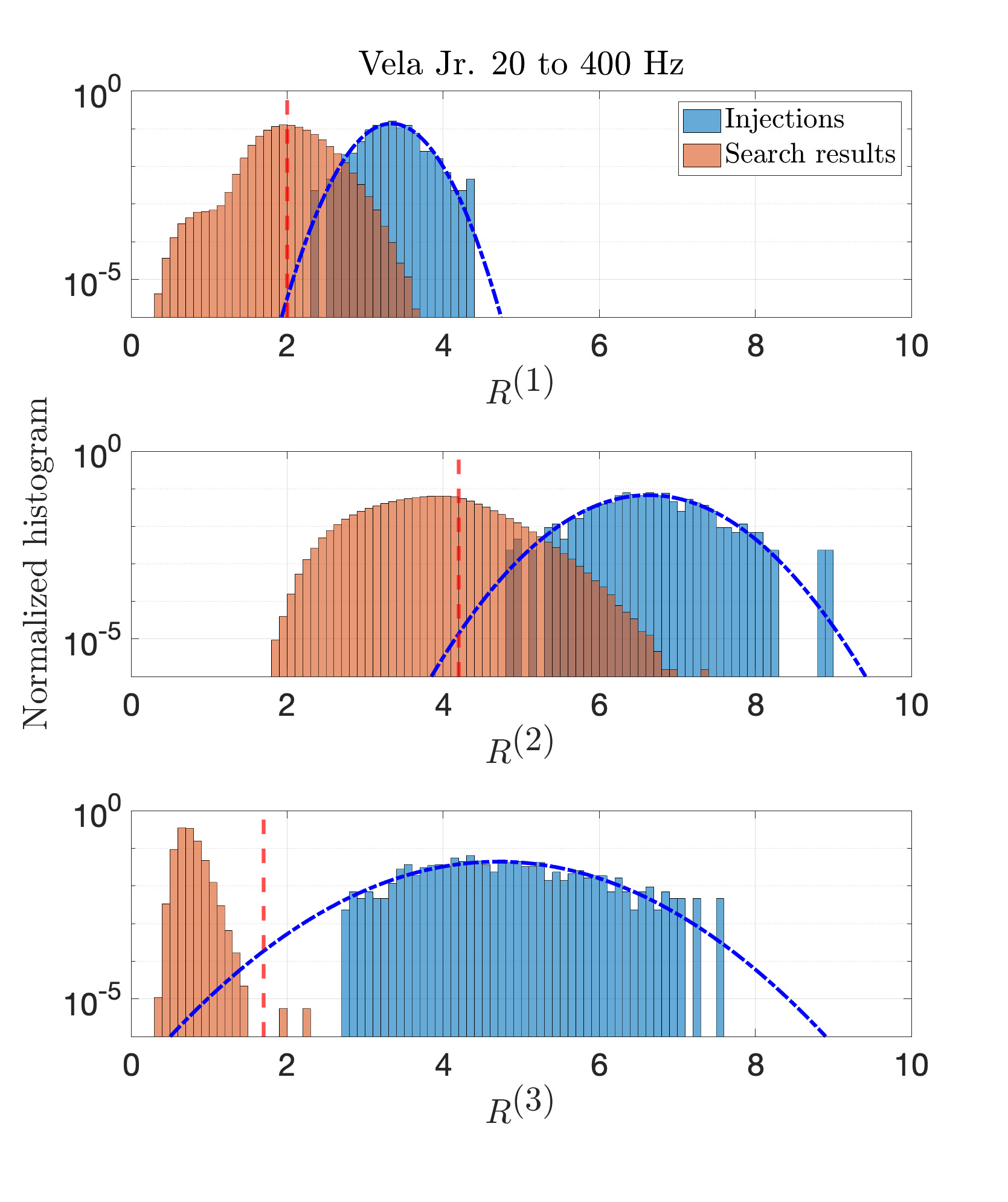}
\caption{$R^{\textrm{(a)}}$ distribution for the Vela Jr. searches below 400 Hz. The orange histograms (left-most) show the distribution of search result candidates. 
The blue histograms (right-most) show the distribution of our target-signal population and its Gaussian fit is the blue dashed line. All result candidates to the left of the red vertical line are discarded.}
\label{fig:veto_V1_2fr_all}
\end{figure}

\begin{figure}[h!tbp]
   \includegraphics[width=\columnwidth]{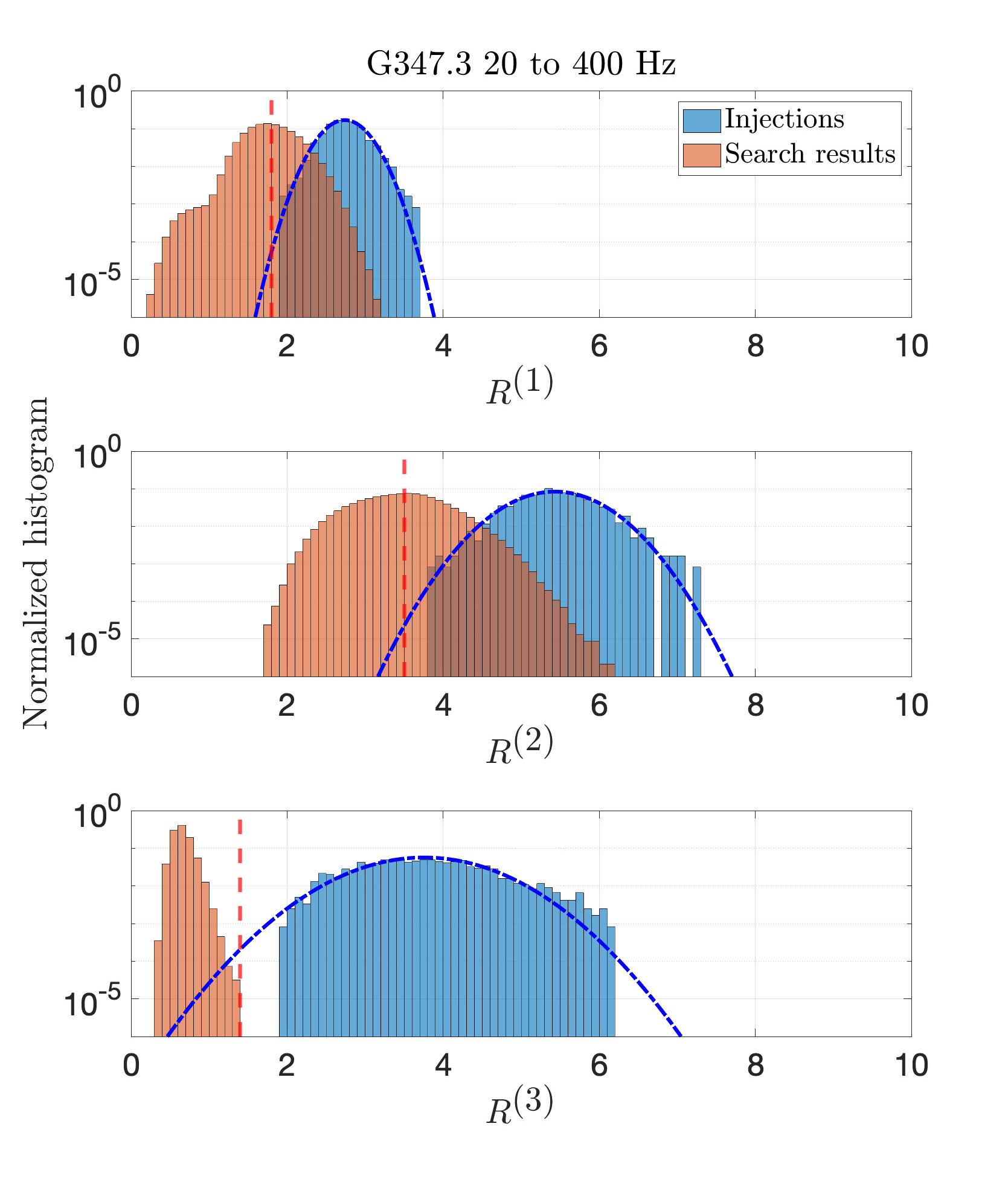}
\caption{$R^{\textrm{(a)}}$ distribution for the G347.3 searches below 400 Hz. The orange histograms (left-most) show the distribution of search result candidates. 
The blue histograms (right-most) show the distribution of our target-signal population and its Gaussian fit is the blue dashed line. All result candidates to the left of the red vertical line are discarded.}
\label{fig:veto_G1_2fr_all}
\end{figure}

\begin{figure}[h!tbp]
   \includegraphics[width=\columnwidth]{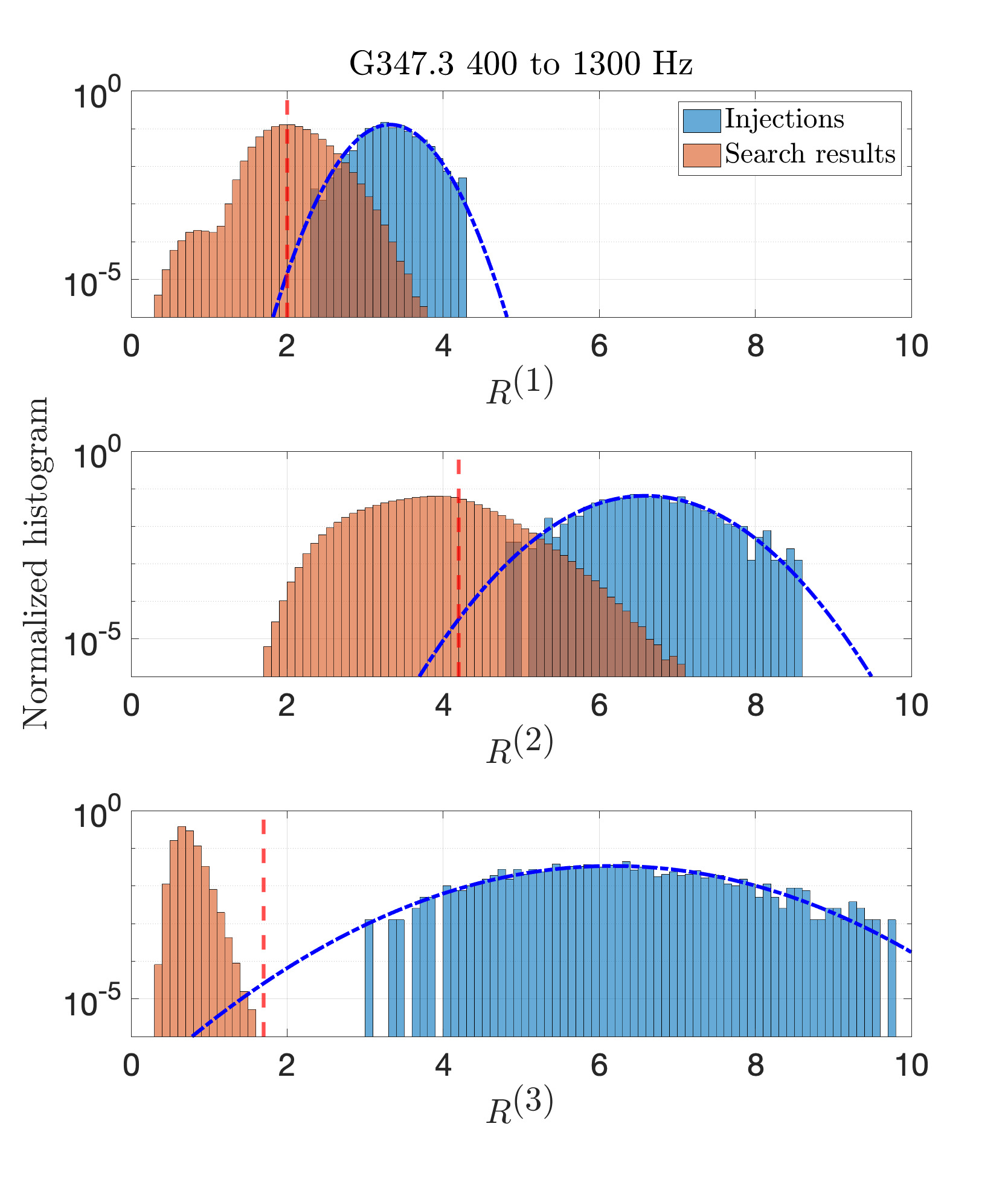}
\caption{$R^{\textrm{(a)}}$ distribution for the G347.3 searches from 400 to 1300 Hz. The orange histograms (left-most) show the distribution of search result candidates. 
The blue histograms (right-most) show the distribution of our target-signal population and its Gaussian fit is the blue dashed line. All result candidates to the left of the red vertical line are discarded.}
\label{fig:veto_G2_2fr_all}
\end{figure}

The threshold values $R^{\textrm{(a)}}_{\textrm{thr}}$ are determined based on the distributions of $R^{\textrm{(a)}}$ for our target signals, which are shown in Figs. \ref{fig:veto_V1_2fr_all}, \ref{fig:veto_G1_2fr_all} and \ref{fig:veto_G2_2fr_all}. In particular, we fit the normalized histograms with a Gaussian and set the threshold such that the false dismissal probability is always $\lambda_{\text{FD}}\ll 0.1\%$, as shown in Table \ref{tab:veto_all}.
We experiment in constructing different discriminators, also using the transient- and line-robust statistic $\BSNtsc$ but find that Eq.~\ref{eq:Rdef} defines the most efficient detection statistic to identify the candidates for the next stage. With ``most efficient'' here we mean that at fixed false dismissal it yields the lowest false alarm.

Before Stage 3, the search boxes for Stage (a+1) search around each candidate are the uncertainties $\pm\Delta f$, $\pm\Delta\fdot$ and $\pm\Delta\fddot$ on the parameters of the candidates recovered at Stage (a). These are determined based on the distribution of distances between the true parameters and the recovered parameters of fake signals  added to the data and searched-for using a Stage (a) search. The uncertainties are such that for $\ge 99\%$ of the considered signal population, the candidates' parameters are closer to the true fake signal parameters than the uncertainty. These are also detailed in Table \ref{tab:cr}. For Stage 3 the search areas are detailed in Section \ref{sec:stage3}. 

For each search, 2000 fake signals are injected into the original data. For each of the low frequency searches of Vela Jr and G347.3, there are 2000 fake signal  randomly distributed in the frequency range from 200 to 220 Hz. The reason for choosing this band is because it is not disturbed. In high frequency search of G347.3, these 2000 fake signals are randomly distributed in the frequency range from 715 to 735 Hz for the same reason. The spin-down values of these fake signals are log-uniformly distributed in their possible spin-down ranges which is described by Eq.s~\ref{eq:Priors}. 
The $\cos\iota$ and $\psi$ parameters  distributed described above are uniformly distributed in their physical ranges: $-1\leq \cos\iota \leq 1$, $-\pi/4 \leq\psi \leq \pi/4$. 

We choose the amplitudes of the fake signals so that they are representative of our target population. We estimate the weakest signal that we can detect based  on the number of  candidates that we select from Stage 0, and take that as an indication of our target amplitudes. It turns out that these target amplitudes are at effective sensitive depths (Eq.31 of \cite{Tenorio_2021}) of about $\approx100 ~[1/\sqrt{\textrm{Hz}}]$ for G347.3 high frequency and Vela Jr. search, and $\approx105~[1/\sqrt{\textrm{Hz}}]$ for  G347.3 low frequency, respectively.

Before we describe the follow-up searches of Stages 1-3, in the next two sections \ref{sec:banding} and \ref{sec:clustering} we explain how we select the  few million candidates that we follow-up, starting from the $3\times 10^{10} ~\mathrm{\EatH}$ results.

\subsection{Banding and identification of disturbed bands}
\label{sec:banding}

The frequency search range is split in 50 mHz bands. All the results for signal frequencies in the same 50 mHz and for signal frequency derivative values in the entire search range are collected together.  We carry out a visual inspection of these results in each band and identify bands grossly affected by disturbances, as described in \citep{LIGOScientific:2017wva,Zhu:2016ghk}. In fact, even after the removal of disturbed data caused by spectral artefacts of known origin, and in spite of using a line-robust detection statistic for ranking the Einstein@Home search results, the statistical properties of the results are not uniform across the search band. It turns out that in the low frequency band 20-400 Hz,  1.1\% of the 50 mHz bands are disturbed, while in the high frequency band 400-1300 Hz, only about 0.6\% of the bands are disturbed. The candidates from these disturbed bands are ignored in our further investigations. A list of the 50 mHz disturbed bands is given \citep{AEIULurl}.

\subsection{Clustering}
\label{sec:clustering}

The mean mismatch of  our searches is about 13\% which means that neighboring templates are not entirely independent.
No matter whether it is a disturbance or a signal, both can generate multiple adjacent detection-statistic results that are prominent enough to make them into our top list. However, we don't need to follow up all of them, because their origin is the same.  To avoid  wasting computing resources on these ``repetitive" results, we use a clustering procedure that identifies results that originate from the same underlying cause and that allows us to consider them as a single candidate for the purpose of follow-ups. Different clustering have been developed over time to process the large Einstein@Home top lists \citep{Singh:2017kss,Beheshtipour:2020zhb,Beheshtipour:2020nko}. Here we consider our most recent and effective one \citep{den_cluster}. We refer to these output of the clustering procedure as ``candidates".

The clustering procedure significantly reduces the number of follow-ups, enabling us to further examine approximately 5 million candidates and achieve the desired search sensitivities.
In Table \ref{tab:HighThresholds}, we list the Stage 0 detection statistic thresholds $\BSGLtLr$ and the number of candidates $\mathrm{N_{S0}}$ above these thresholds after the clustering procedure. More parameters are actually necessary to specify the clustering procedure and these are given in the Appendix~\ref{App:clustering} for the reader interested in this somewhat technical information. 

\begin{table*}[ht]
\centering
\begin{tabular}{|c|c|c|c|}
\hline
\hline
 \TBstrut&Vela Jr. (20 - 400 Hz) & G347.3 (20 - 400 Hz) & G347.3 (400 - 1300 Hz) \\
\hline
\hline
 \TBstrut$~\BSGLtLr~$ &  $\ThrVelalow$ & $\ThrGlow$  & $\ThrGmid$ \\
 \hline
\TBstrut$\mathrm{N_{S0}}~$ &  $\NumVelalow$ & $\NumGlow$  & $\NumGmid$ \\
 \hline
\hline  

\end{tabular}
\caption{Stage 0 detection statistic thresholds and the number of candidates for follow-ups.}
\label{tab:HighThresholds}
\end{table*}

\begin{figure}[h!]
    \includegraphics[width=1 \columnwidth]{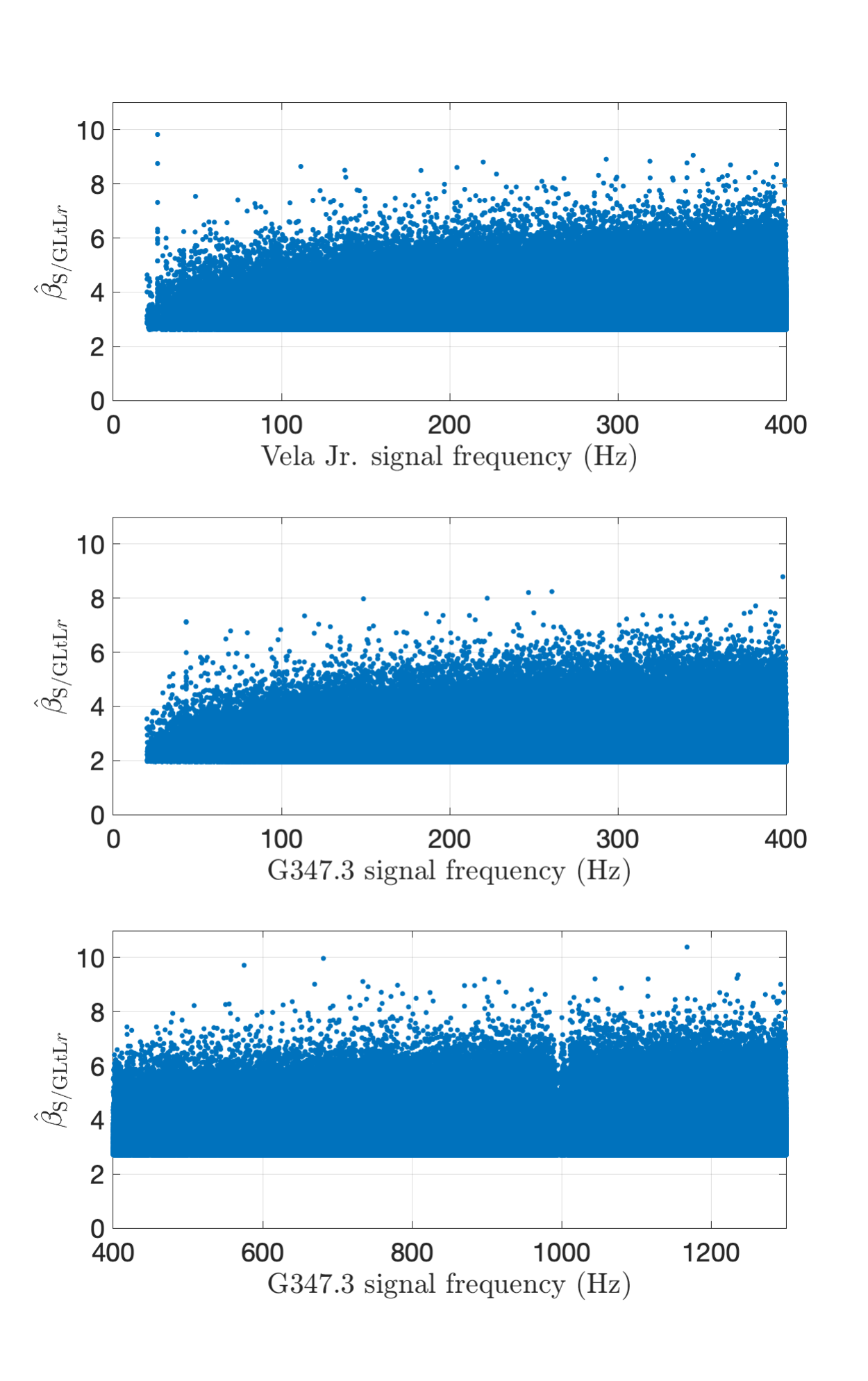}
    \caption{Detection statistic $\BSGLtLr$ as a function of signal frequency for the candidates surviving Stage 0. There are \NumVelalow~such candidates for Vela Jr., $\NumGlow$ candidates for G347.3 below 400 Hz and $\NumGmid$ candidates for G347.3 above 400 Hz. Note that for G347.3 at around 1000 Hz, there is a ``dent" which indicates that the number of candidates from that frequency region is significantly smaller than the  other frequency region. 
The cause of this is the large difference of power spectral densities between two detectors, amounting on average to $\approx 10$, which causes lower $\BSNtsc$ values than if the detectors had comparable sensitivities. This is a known effect already highlighted by \citep{Keitel_2015}.}
\label{fig:CandsStage0}
\end{figure}

\begin{table*}[ht]
\centering
\begin{tabular}{|c|c|c|c|}
\hline
\hline
 search set-up &  Stage 1 (O2)  & Stage 2 (O2) & Stage 3 (O3a) \\
\hline
\hline
 \TBstrut$\Tcoh$ & 2760 hrs & 5496 hrs (whole span of O2) & 1440 hrs \\ 
 \hline
\TBstrut$\Nseg$ & 2 & 1 & 3\\
  \hline
\TBstrut$\delta f$ & $3.5 \times 10^{-8}$ Hz & $9.7 \times 10^{-9}$ Hz  & $6.7\times 10^{-8}$ Hz \\
 \hline
 \TBstrut$\delta {\dot{f_c}}$ &  $2.4\times 10^{-14}$ Hz/s  & $2.9\times 10^{-15}$ Hz/s  & $8.7\times 10^{-14}$ Hz/s  \\
  \hline
\TBstrut$\gamma_1$ & 11 & 1 & 7 \\
  \hline
\TBstrut$\delta {\ddot{f_c}}$ & $ 3.6\times 10^{-21} ~{\textrm{Hz/s}}^2 $ & $ 1.5\times 10^{-21} ~{\textrm{Hz/s}}^2 $  & $ 8.1\times 10^{-21} ~{\textrm{Hz/s}}^2 $ \\
  \hline
\TBstrut$\gamma_2$ & 3 & 1 & 5 \\
 \hline
\TBstrut$\avMis$ & 7.6\% &  2.0\%  &  5.4\% \\
 \hline
\hline  
\end{tabular}
\caption{Spacings on the signal parameters used for the templates in stages of follow-up searches and the average mismatches.}
\label{tab:GridSpacingsall}
\end{table*}

\begin{table*}[ht]
\centering
\begin{tabular}{|c|c|c|c|c|c|c|c|c|c|c|c|c|c|c|}
\hline
\hline
 & \multicolumn{4}{c|} {Vela Jr. (20 - 400 Hz)} & \multicolumn{4}{c|} {G347.3 (20 - 400 Hz)} &  \multicolumn{4}{c|} {G347.3 (400 - 1300 Hz)} \\
\hline 
\hline
&$ R^{\textrm{(a)}}_{\textrm{thr}}$ & $\lambda_{\text{FD}}$ &$N_\textrm{in}$ & $N_\textrm{out}$ &$ R^{\textrm{(a)}}_{\textrm{thr}}$ &$\lambda_{\text{FD}}$ & $N_\textrm{in}$ & $N_\textrm{out}$ &$ R^{\textrm{(a)}}_{\textrm{thr}}$ & $\lambda_{\text{FD}}$ &$N_\textrm{in}$ & $N_\textrm{out}$\\
\hline
Stage 1 &2.0&$1.9\times10^{-6}$&1\,200\,000&644\,075&1.8&$3.0\times10^{-5}$&1\,000\,000&464\,034&2.0&$1.0\times10^{-5}$&3\,100\,000&1\,427\,335 \\
\hline
Stage 2 &4.2&$1.9\times10^{-5}$&644\,075&182\,073&3.5&$2.5\times10^{-5}$&464\,034&220\,124&4.2&$5.1\times10^{-5}$&1427\,335&381\,588 \\
\hline
Stage 3 &1.7&$4.7\times10^{-4}$&182\,073&0&1.4&$4.0\times10^{-4}$&220\,124&0&1.7&$8.0\times10^{-5}$&381\,588&0 \\
\hline 
\hline
\end{tabular}
\caption{Candidates vetoing overview of follow-up searches.}
\label{tab:veto_all}
\end{table*}

\begin{table*}[ht]

\centering

\begin{tabular}{|c|c|c|c|c|c|c|c|c|c|c|c|}
\hline
\hline
 & \multicolumn{3}{c|} {Vela Jr. (20 - 400 Hz)} & \multicolumn{3}{c|} {G347.3 (20 - 400 Hz)} &  \multicolumn{3}{c|} {G347.3 (400 - 1300 Hz)} \\
\hline 
\hline
@$\tau_{\textrm{SSB}}^{\textrm{O2}}$ &$\Delta f$ (Hz/s)& $\Delta\fdot$ (Hz/s) &$\Delta\fddot$ (${\textrm{Hz/s}}^2$) &$\Delta f$ (Hz/s)& $\Delta\fdot$ (Hz/s) &$\Delta\fddot$ (${\textrm{Hz/s}}^2$)&$\Delta f$ (Hz/s)& $\Delta\fdot$ (Hz/s) &$\Delta\fddot$ (${\textrm{Hz/s}}^2$) \\ 
\hline
Stage 0 &$4.0\times10^{-7}$  & $4.0\times10^{-14}$  & $4.0\times10^{-20}$ 
&$3.0\times10^{-7}$  & $2.5\times10^{-14}$  & $1.5\times10^{-20}$ 
&$4.0\times10^{-7}$  & $5.0\times10^{-14}$  & $4.0\times10^{-20}$ \\
\hline
Stage 1 &$1.0\times10^{-7}$  & $1.0\times10^{-14}$  & $6.5\times10^{-21}$ 
&$1.0\times10^{-7}$  & $1.0\times10^{-14}$  & $5.0\times10^{-21}$ 
&$6.0\times10^{-8}$  & $9.0\times10^{-15}$  & $6.0\times10^{-21}$ \\
\hline
Stage 2 &$7.0\times10^{-8}$  & $4.0\times10^{-15}$  & $3.5\times10^{-21}$ 
&$4.0\times10^{-8}$  & $5.0\times10^{-15}$  & $3.3\times10^{-21}$ 
&$3.0\times10^{-8}$  & $5.0\times10^{-15}$  & $3.0\times10^{-21}$ \\
\hline
\hline

\hline
@$\tau_{\textrm{SSB}}^{\textrm{O3a}}$ &$\Delta f$ (Hz/s)& $\Delta\fdot$ (Hz/s) &$\Delta\fddot$ (${\textrm{Hz/s}}^2$) &$\Delta f$ (Hz/s)& $\Delta\fdot$ (Hz/s) &$\Delta\fddot$ (${\textrm{Hz/s}}^2$)&$\Delta f$ (Hz/s)& $\Delta\fdot$ (Hz/s) &$\Delta\fddot$ (${\textrm{Hz/s}}^2$) \\ 
\hline
{Stage 2} &$8.5\times10^{-6}$  & $2.4\times10^{-13}$  & $3.5\times10^{-21}$ 
&$8.1\times10^{-6}$  & $2.3\times10^{-13}$  & $3.3\times10^{-21}$ 
&$7.4\times10^{-6}$  & $2.1\times10^{-13}$  & $3.0\times10^{-21}$ \\
\hline

\end{tabular}
\caption{The uncertainties $\Delta f$, $\Delta\fdot$ and $\Delta\fddot$ on the parameters of the candidates recovered at Stage 0, 1 and 2. The last row shows the uncertainties translated to the Stage 3 follow-up reference time ($\tau_{\textrm{SSB}}^{\textrm{O3a}}$). These are much larger than the uncertainties in previous stages as explained in Section~\ref{sec:stage3}. }
\label{tab:cr}
\end{table*}

\subsection{Stage 1}
\label{sec:stage1}

Fig. \ref{fig:CandsStage0} is a scatter plot showing the frequency and detection-statistic value of the candidates that are followed-up in Stage 1.  The maximum recorded detection statistic increases with frequency because the frequency-derivative ranges increase with frequency and hence the number of searched templates is larger (trials factor effect). Also the number of candidates increases with frequency.

\begin{figure}[h!tbp]
  \includegraphics[width=\columnwidth]{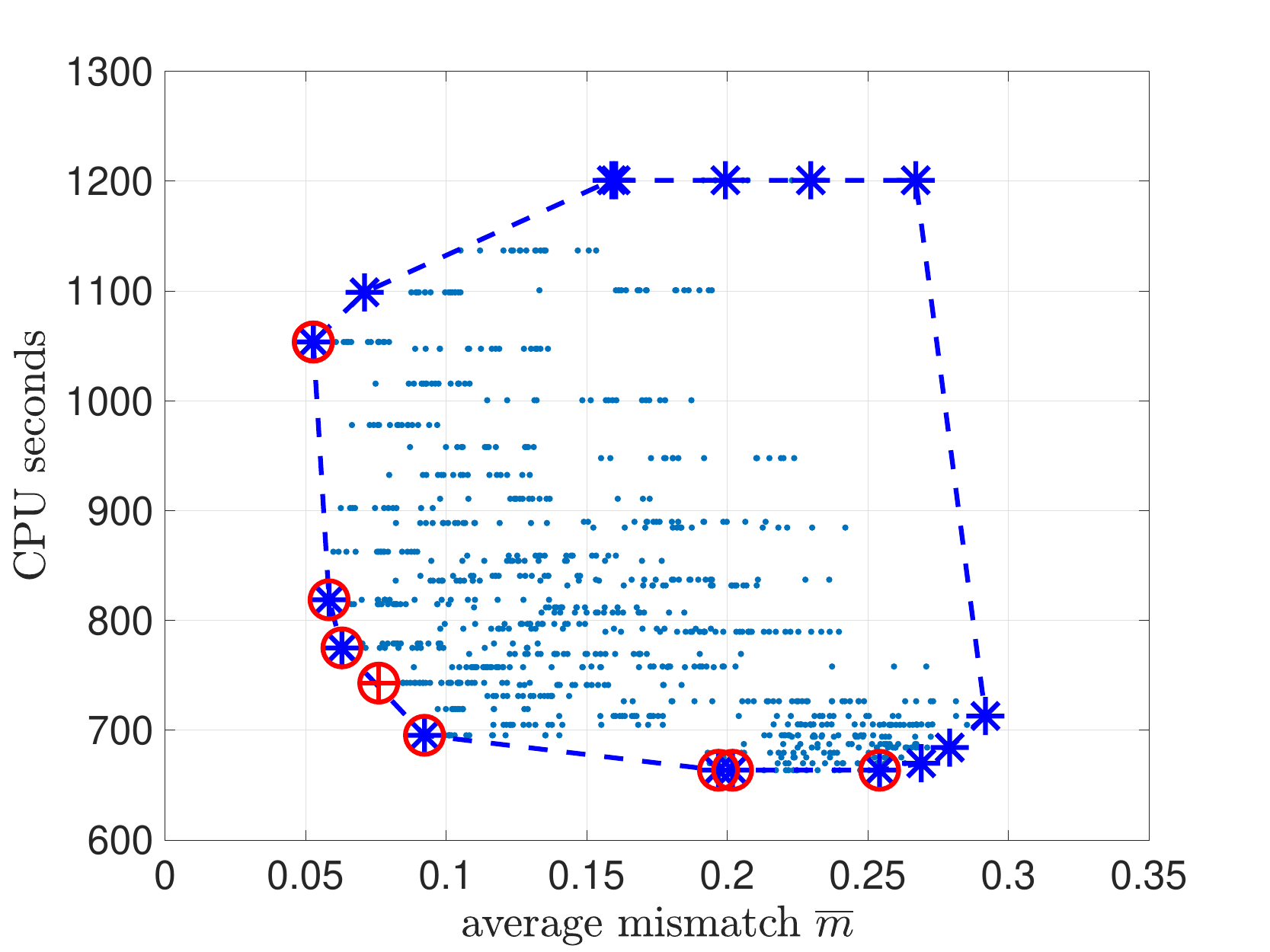}
\caption{Computing cost per candidate of 920 putative Stage 1 set-ups with $\Tcoh=2670$ hrs, for the Vela Jr. search. The 8 red circles are the lowest computing cost set-ups at a given mismatch. The thick circle with a red `+' is the chosen Stage 1 set-up.}

\label{fig:G1_FU1_cost}
\end{figure}

As shown in Table \ref{tab:GridSpacingsall},  Stage 1 uses the same coherent time-baseline $\Tcoh=2760$ hrs for all targets. 
A mismatch of $\lesssim 8\%$ was chosen as it allows to perform the follow-up in a few days for each target, using a few thousand nodes of the ATLAS computing cluster \citep{ATLAS}. Fig. \ref{fig:G1_FU1_cost} shows the chosen set-ups over all the ones that were tested for the Vela Jr. Stage1 search, and illustrates the optimization strategy.

We use the following thresholds on $R^{(1)}$ for Vela Jr.,  G347.3 below 400 Hz and above 400 Hz respectively: \RoneVlowFst, ~\RoneGlowFst ~and \RoneGmidFst. Approximately \NrejectFUoneVlowFst, ~ \NrejectFUoneGlowFst ~and \NrejectFUoneGmidFst ~candidates survive to the next stage, corresponding to 53\%, 47\% and 46\% of the Stage 1 candidates for Vela Jr., G347.3 below 400 Hz and G347.3 above 400 Hz respectively.

\subsection{Stage 2}
\label{sec:stage2}

In this stage we follow up the candidates that survive Stage 1. We use a fully coherent search over the entire O2 data set. The search set-up has an average mismatch of 2.0\% and the details are shown in Table \ref{tab:GridSpacingsall}.

A few $\times 10^5$ candidates survive this stage for each target, overall corresponding to about 15\% of the initial Stage 0 ones. The exact values are given in Table \ref{tab:veto_all}, together with the $R^{\textrm{(2)}}$ thresholds used.

\subsection{Stage 3}
\label{sec:stage3}

The $7.8\times10^{5}$ candidates surviving the Stage 1 and 2 follow-ups are now investigated using a new data set, O3a. This is comparable in duration with the O2 set and spans about 180 days. 
The O3a data is more sensitive compared to O2 and the duty factor is higher, but due to the shorter $\Tcoh$ of Stage 3 the expected detection statistic values are on average lower than those of Stage 2, but still higher than those of Stage 0. The critical difference though is that the data is different and hence the noise distribution now distinctly separates from the signal distribution and this allows to veto most remaining candidates.

The O2 candidate parameters refer to a reference time $\tau_{\textrm{SSB}}^{\textrm{O2}}$ at roughly the mid time of the O2 run. Now using O3a data, it is most convenient to evolve the candidate parameters to a reference time that is close to the middle of the O3a run, $\tau_{\textrm{SSB}}^{\textrm{O3a}}$. The distance $\Delta\tau \approx$ 800 days:
\begin{equation}
\label{eq:fevolve}
\begin{cases}
f(\tau_{\textrm{SSB}}^{\textrm{O3a}})=f(\tau_{\textrm{SSB}}^{\textrm{O2}}) + \dot{f}(\tau_{\textrm{SSB}}^{\textrm{O2}}) \Delta\tau + \ddot{f}(\tau_{\textrm{SSB}}^{\textrm{O2}}) \frac{\Delta\tau^2}{2} \\
\dot{f}(\tau_{\textrm{SSB}}^{\textrm{O3a}})=\dot{f}(\tau_{\textrm{SSB}}^{\textrm{O2}}) + \ddot{f}(\tau_{\textrm{SSB}}^{\textrm{O2}})\Delta\tau \\
\ddot{f}(\tau_{\textrm{SSB}}^{\textrm{O3a}})= \ddot{f}(\tau_{\textrm{SSB}}^{\textrm{O2}})
\end{cases}
\end{equation}
The search regions of the O3a search are the candidate parameter uncertainties from the previous stage, augmented due to the difference in reference time. In particular,  
the $\Delta\dot{f}$ and $\Delta{\ddot{f}}$  of the Stage 2 candidates give rise to larger uncertainties in $\fdot$ and $f$ on data after the 800-day gap. The last row of Table~\ref{tab:cr} lists the extent of the search regions to the right and left of the nominal candidate parameters. 
All in all the number of templates of each candidate follow-up in O3a varies between $3.9\times10^{5}$ and $5.3\times10^{5}$ across the different targets and the different frequency bands. 

The bottom subplots in Figs. \ref{fig:veto_V1_2fr_all}, \ref{fig:veto_G1_2fr_all} and \ref{fig:veto_G2_2fr_all} show the $R^{\textrm{(3)}}$ distributions for the data and for the fake signal population. The bulk of the blue histogram of the bottom subplot in Fig. \ref{fig:veto_G1_2fr_all}  for the low-frequency G347.3 target is at lower values than the other two, due to the fact that the original Stage 0 search for that search has a higher $\Tcoh=1080$ hrs than the other two. The $R^{\textrm{(3)}}$ distribution from the high-frequency G347.3 search extends to higher values than the Vela Jr. distribution because the improvement in sensitivity of the O3a data is greater at high frequencies than at low frequencies. The power spectral density ratio of O3a to O2 at 200 Hz is 0.45, while at 730 Hz, this ratio is 0.31.

Only two candidates survive the Stage 3 cut, from the Vela Jr. search, as seen in Fig. \ref{fig:veto_V1_2fr_all}.  
One is at a frequency $ \approx$ 42 Hz with a  ${2\F_r}=37$ and the other is at a frequency $ \approx$ 100 Hz with a  ${2\F_r}=30$.
However the corresponding $\BSGLtLr$ values are $-18$ and $-23$ respectively, making them the two lowest values of all 180,
 000 Vela Jr. candidates followed-up in Stage 3, suggesting that a disturbance in one of the two detectors is the likely cause of the elevated values of $2\F_r$. This is confirmed by inspecting the average power spectral density from the O3a data for the detectors. We find large spectral features right around the candidates' frequencies in the H1 detector.

\section{Results}
\label{sec:results}

\subsection{Upper limits on the gravitational wave amplitude}

Since none of the candidates that were investigated can be associated with a signal, we determine frequentist 90\% confidence upper limits on the maximum gravitational wave amplitude consistent with this null result in every half Hz band, $h_0^{90\%}(f)$.  We use the same procedure as in our previous search \citep{Ming_2022}, and describe it using the same text since it is the clearest way to do so. 

The $h_0^{90\%}(f)$ is the GW amplitude such that 90\% of a population of signals with parameter values in our search range would have survived the Stage 0 of our hierarchical search. Since the following stages have a negligible false dismissal rate for the same signal population, this means that the signal would have survived as a candidate up to the last stage.

We now recall the upper limit procedure and use our same text as \citep{Ming_2022}, since we believe that was a very clear explanation. We use quotes to indicate that text. 

``In each half Hz band, 200 simulated signals with a constant intrinsic amplitude value of $h_0$ are injected into the real detector data. Subsequently, the data undergoes processing as if it were the data being searched, including gating and line cleaning. 

The parameters of simulated signals, the frequency, inclination angle $\cos\iota$, polarization $\psi$ and initial phase values, are uniformly randomly distributed in their respective ranges. The spin-down values, $\dot{f}$ and $\ddot{f}$, are log-uniformly randomly distributed in their respective ranges. 

A search is conducted to recover each injection using the same search set-up as the original Einstein@Home search uses. However, to conserve computational resources, the search is constrained and focuses on the parameter space surrounding the simulated signal." For an injected signal to be considered as recovered it should survive Stage 0. 
If an injection falls  in a 50 mHz band which was marked as disturbed, this injection are marked as unrecovered.

``We repeat this procedure for various values of $h_0$. 
For every value of $h_0$,  the fraction of detected injections is accounted and yields a detection efficiency, or detection confidence,  $C(h_0)$. The $h_0$ versus confidence $C(h_0)$ data is fit with a sigmoid of the form: 
\begin{equation}
C(h_0)={1\over{1+\exp({{{\textrm{a}}-h_0}\over{\textrm{b}}})}},
\label{eq:sigmoidFit}
\end{equation}
and from it  the $h_0$ amplitude that corresponds to 90\% confidence is read-off as our upper limit value $h_0^{90\%}$.

\begin{figure}[h!tbp]
   \includegraphics[width=\columnwidth]{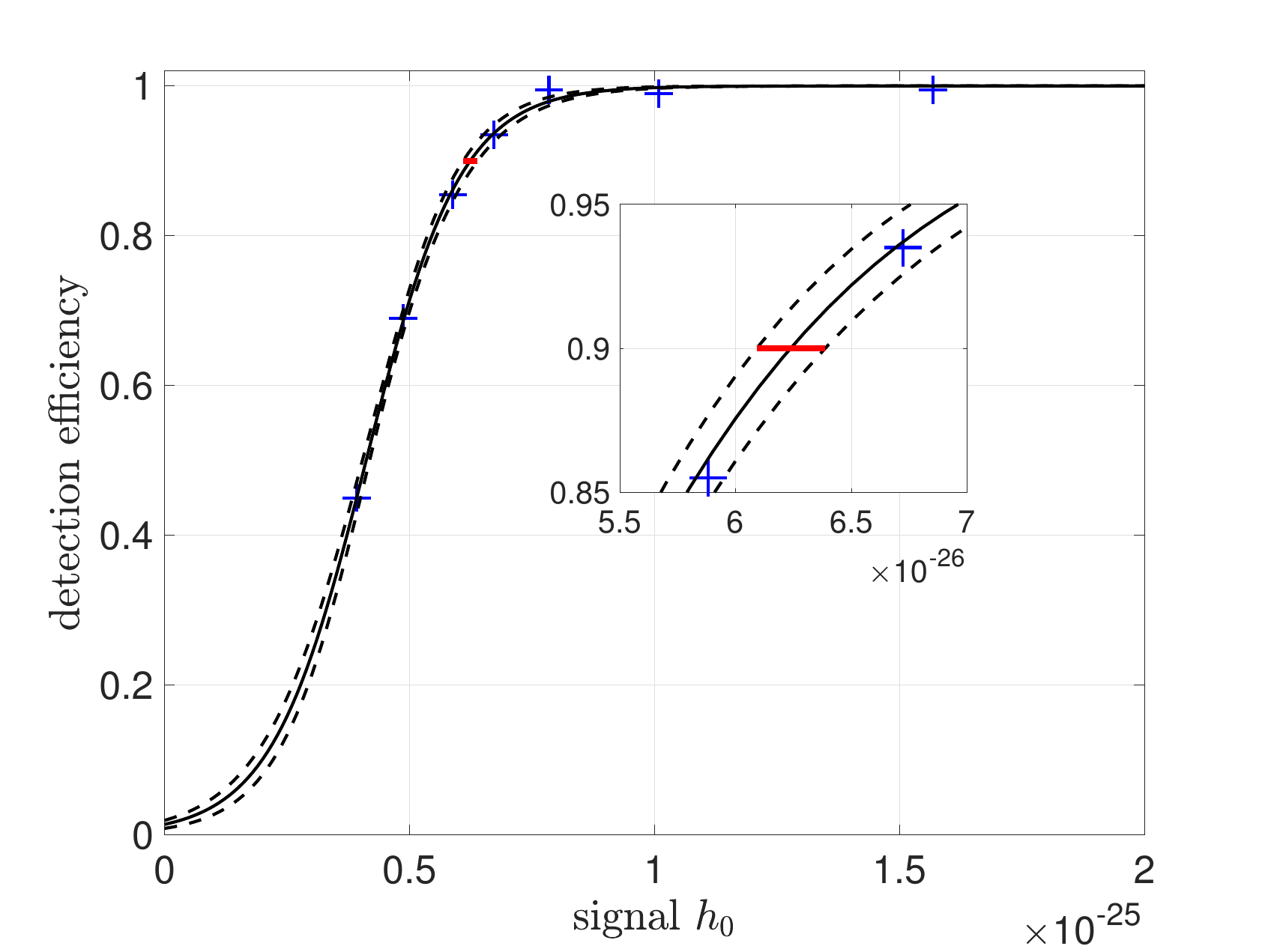}
\caption{Blue crosses: measured detection efficiency $C(h_0)$ from the G347.3 signal search-and-recovery Monte Carlos with signal frequencies between 161.5 and 162 Hz. The solid line is the best fit and the dashed lines represent 95\% confidence intervals on the fit. The red line marks the 90\% detection confidence level, with the uncertainties introduced by this fitting procedure  is $\le5\%$. The inset shows a zoom around the 90\% confidence level.}
\label{fig:exampleSigmoid}
\end{figure}

We utilise the Matlab nonlinear regression routine {\tt{nlpredci}} to yield the best-fit for ${\textrm{a}}$ and ${\textrm{b}}$ values and the covariance matrix.  
This  covariance matrix is then used to compute the 95\% credible interval on the fit of $h_0^{90\%}$.
In Fig. \ref{fig:exampleSigmoid}, an illustrative example of the sigmoid curve fitting is demonstrated for the frequency band of 161.5-162 Hz.
The best fit value for $h_0^{90\%}$ in this particular band is $6.2\times10^{-26}$. The uncertainties introduced by the procedure amount is less than 5\%.

The overall uncertainty in the upper limit consists of the uncertainty resulting from the fitting procedure as well as the calibration uncertainties. As a conservative estimate, we adopt a calibration uncertainty of 5\% based on \citep{PhysRevD.96.102001}.''

\begin{figure}[h!]
   \includegraphics[width=1 \columnwidth]{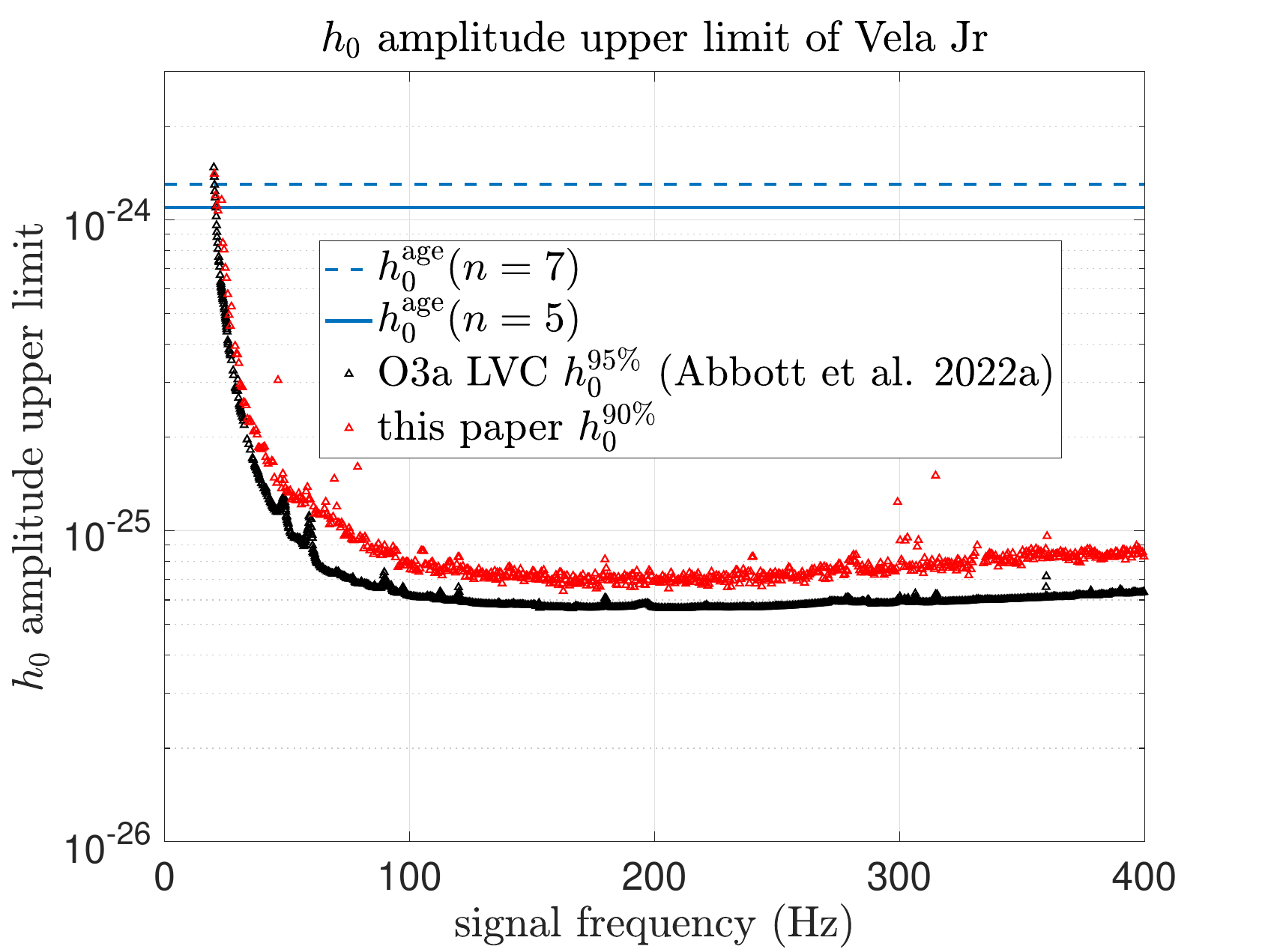}
\caption{90\% confidence upper limits on the gravitational wave amplitude of continuous gravitational wave signals from Vela Jr. for signals with frequency between  20 to 400 Hz. 
The red triangles are the results of this search and we compare them with results from the LVC search of the O3a \citep{ligo_o3a_c_v}. 
The dashed blue line at the top shows the age-based upper limit assuming braking index $n=7$ and the solid blue line shows the case assuming $n=5$. We show the most constraining age limit for Vela Jr., i.e. the one assuming the object is farther away (900 pc) and older (5100 years). The limit under the assumption that Vela Jr. is young (700 years) and close-by (200 pc) is $1.7\times 10^{-23}$ for $n=7$ and $1.4\times 10^{-23}$ for $n=5$.
}  
\label{fig:ULs_V1}
\end{figure}

\begin{figure*}
   \includegraphics[width=\textwidth]{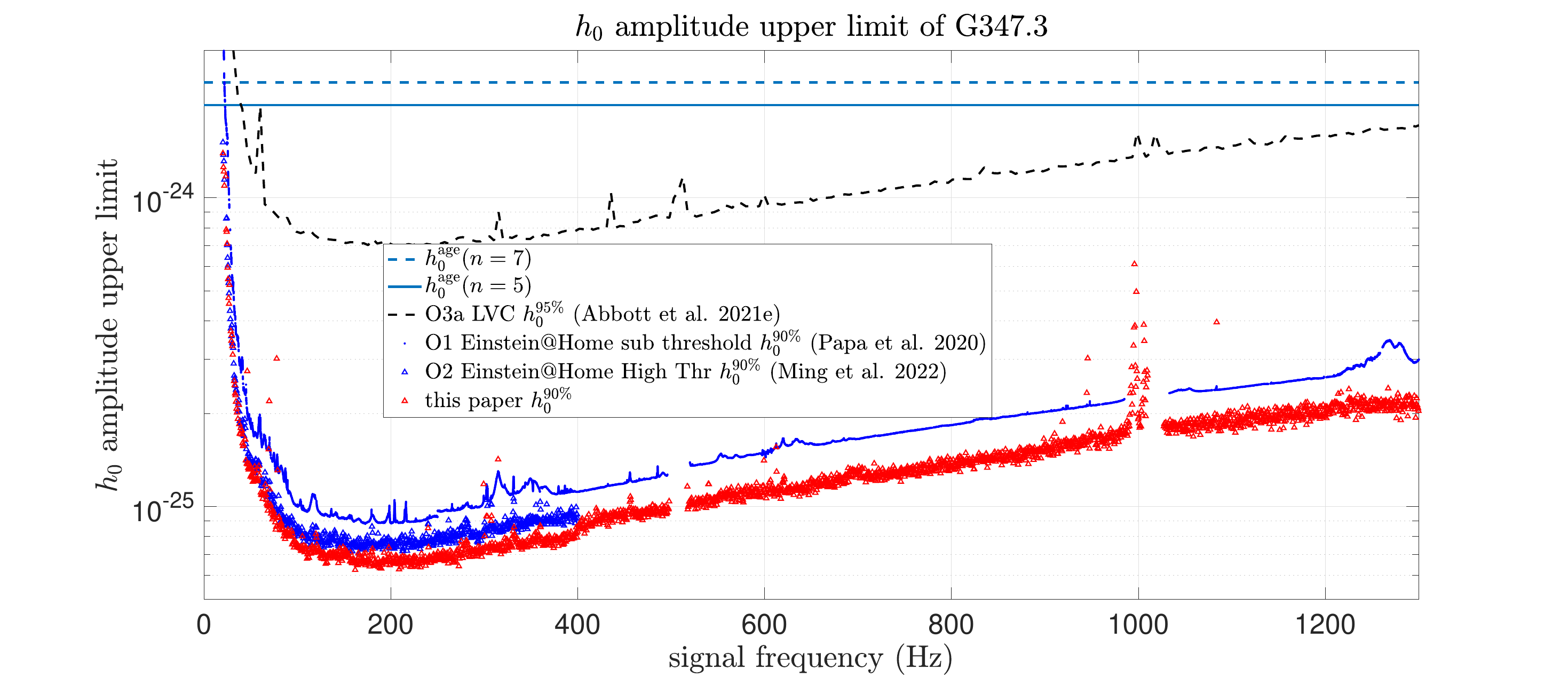}
\caption{90\% confidence upper limits on the gravitational wave amplitude of continuous gravitational wave signals from G347.3 for signals with frequency between  20 to 1300 Hz. 
The lower red triangles are the  results of this search and we compare them with results from previous searches: The black dashed line is the upper limits from the LVC search of the O3a \citep{LIGOScientific:2021mwx};  The blue triangles are Einstein@Home high threshold search results from O2 data \citep{Ming_2022} and blue dots are the  Einstein@Home sub-threshold search results from O1 data \citep{Papa_2020midth}.  The dashed blue line at the top shows the age-based upper limit assuming braking index $n=7$ and the solid blue line shows the case assuming $n=5$. 
}  
\label{fig:ULs_G1G2}
\end{figure*}

The $h_0^{90\%}$ upper limits for Vela Jr. and G347.3 are shown in Fig. \ref{fig:ULs_V1} and Fig. \ref{fig:ULs_G1G2} respectively. We also provide them in machine-readable format at \cite{AEIULurl}. There are bands for which there are no upper limits. This is due to either there being 50 mHz bands marked as ``disturbed" (and hence disregarded) in a given half-Hz band (see Section \ref{sec:banding})  or due to the cleaning procedure, that has removed too much data. The cleaning procedure substitutes disturbed frequency-domain data with Gaussian noise in order to avoid further spectral contamination from ``leakage" in the search results. Those bands are consistently cleaned in the upper-limit Monte Carlos after a signal is injected, so it may happen that most of the injected signal is  removed. When that happens, no matter how loud the signal is, the detection efficiency does not increase. In these bands the 90\% detection rate level cannot be reached and we do not set any upper limit. This reflects the fact that, even if we had a signal there, because of the cleaning procedure, we could not detect it.

For the Vela Jr.  search, we do not set an upper limit in 28 half-Hz bands; correspondingly in the upper limit files we have 732 entries rather than 760. 
For the G347.3 20-400 Hz search we do not set an upper limit in 25 half-Hz bands and correspondingly in the upper limit files we have 735 entries rather than 760. For the G347.3. 400-1300 Hz search  we do not set an upper limit in 89 half-Hz bands and correspondingly in the upper limit files we have 1726 entries rather than 1800. More bands are affected in this range due to the vibrational modes of suspension silica fibre at 500 Hz and its harmonics at around 1000 Hz \citep{TheLIGOScientific:2016agk}.

The Vela Jr. upper limits are shown in Fig.~\ref{fig:ULs_V1}. The most constraining upper limit is at \lowestULvOnefreq~Hz  and measures  \lowestULvOne. Fig. \ref{fig:ULs_G1G2} shows the G347.3 upper limits, with the most constraining upper limit of \lowestULgOne at \lowestULgOnefreq~Hz.

\subsubsection{Sensitivity Depth}

For each of the targets and half-Hz bands we determine the sensitivity depth ${{\mathcal{D}}}^{90\%}$ \citep{GalacticCenterMethod, Dreissigacker:2018afk} of the search corresponding to $h_0^{90\%}(f)$: 
\begin{equation}
{{\mathcal{D}}}^{90\%}:={\sqrt{S_h(f)}\over {h_0^{90\%}(f) }}~~[ {1/\sqrt{\text{Hz}}} ],
\label{eq:sensDepth}
\end{equation}
where $\sqrt{S_h(f)}$ is the noise level associated with a signal of frequency $f$. This quantity is approximately independent of frequency and is useful to characterise the performance of a search on a given data-set. 

For the searches presented here the average values across the frequency ranges are
\begin{equation}
\label{eq:sensDepthResults}
\begin{cases}
\textrm{Vela Jr~~ 20-400 Hz}: ~&{{\mathcal{D}}}^{90\%}\approx 103~[{1/\sqrt{\text{Hz}}} ] \\
\textrm{G347.3~~ 20-400 Hz}: ~&{{\mathcal{D}}}^{90\%}\approx 108~[{1/\sqrt{\text{Hz}}} ] \\
\textrm{G347.3~~ 400-1300 Hz}: ~&{{\mathcal{D}}}^{90\%}\approx 100~[{1/\sqrt{\text{Hz}}} ] 
\end{cases}
\end{equation}

\subsection{Upper limits on astrophysical parameters}

\begin{figure*}[h!tbp]
   \includegraphics[width=\textwidth]{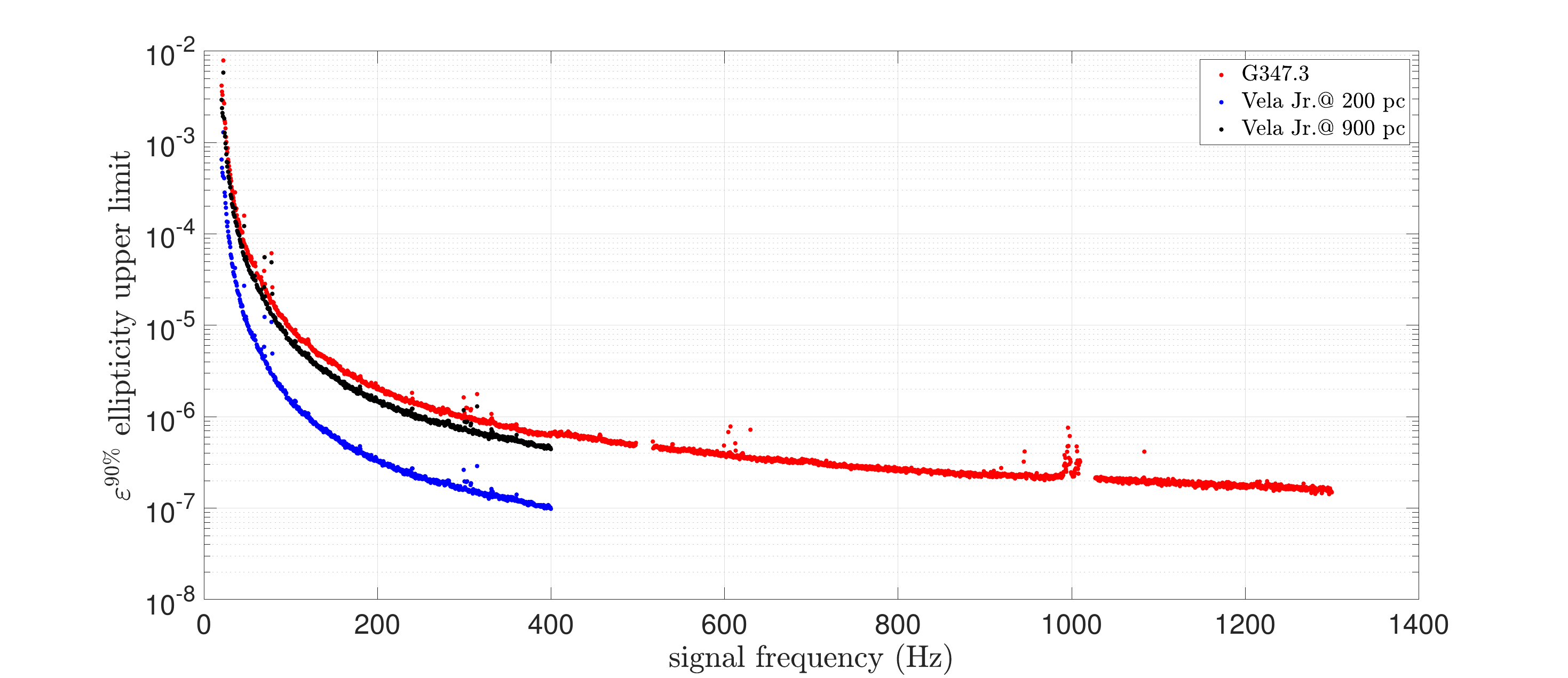}
\caption{ Upper limits on the ellipticity of the Vela Jr. and G347.3 . For Vela Jr. we show two curves, corresponding to two distance estimates: 200 pc and 900 pc.  For G347.3, we assume 1300 pc.}
\label{fig:epsilonULs}
\end{figure*}

The equatorial ellipticity $\varepsilon$ necessary to support continuous gravitational emission with amplitude $h_0$ at a distance $D$ from the source and at frequency $f$ is \citep{zimmermann:1979}
\begin{equation}
\varepsilon = {{c^4}\over {4\pi^2 G}}{{h_0 D}\over {I f^2}}
\label{eq:epsilon}
\end{equation}
where $c$ is the speed of light, $G$ is the gravitational constant and $I$ the principal moment of inertia of the star. Based on this last equation, we can convert the upper limits on the gravitational wave amplitude in upper limits on the ellipticity of the source. The results are shown in Fig. \ref{fig:epsilonULs} assuming a fiducial value of the principal moment of inertia of $10^{38} \textrm{kg m}^2$ and distance estimates for our targets from the literature. 

\begin{figure*}[h!tbp]
   \includegraphics[width=\textwidth]{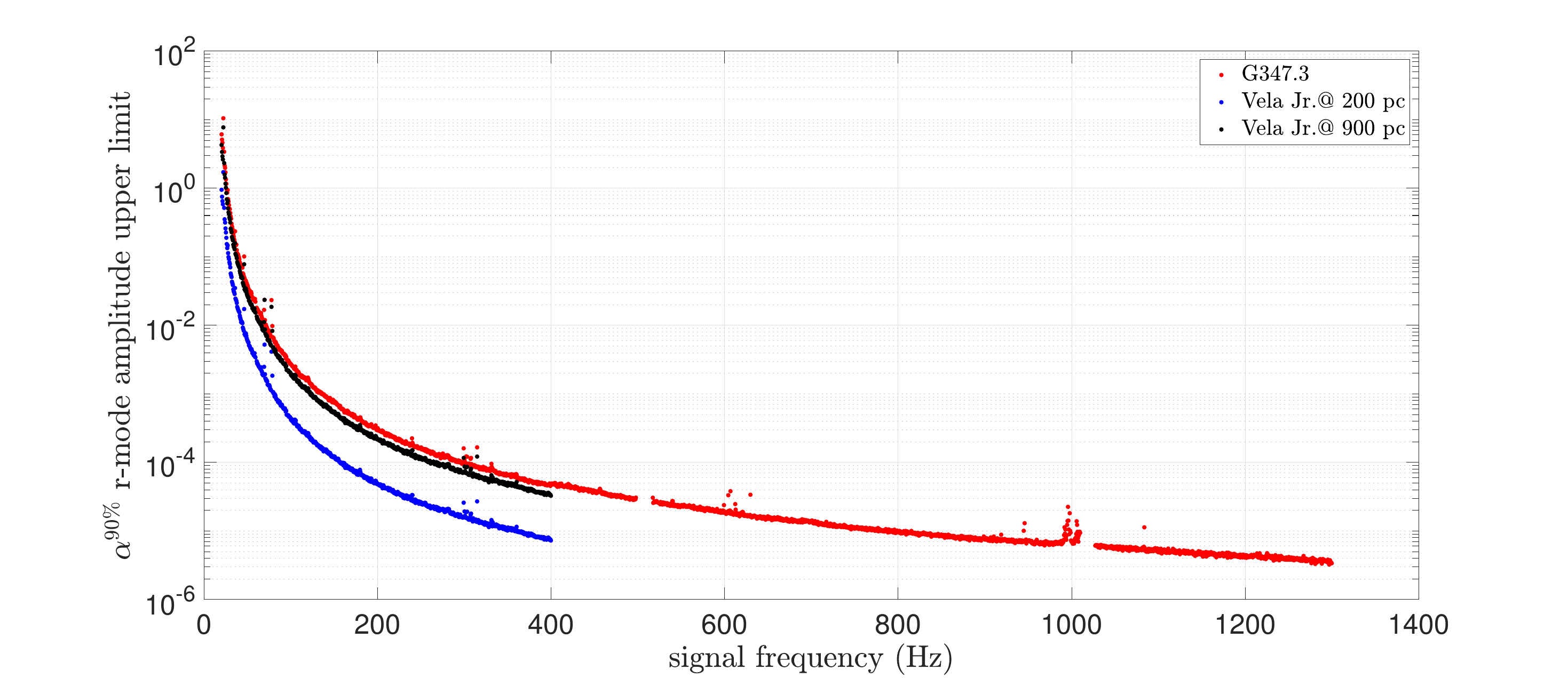}
\caption{Upper limits on the r-mode amplitude of the two targets. For Vela Jr. we show two curves, corresponding to two distance estimates: 200 pc and 900 pc.  For G347.3, we assume 1300 pc.}  
\label{fig:alphaULs}
\end{figure*}

Assuming a 1.4 M$_\odot$ neutron star, with a radius of $11.7$ km, and the same value for $I$ as used above, the r-mode amplitude $\alpha$ that would support continuous gravitational wave emission with amplitude $h_0$ at a frequency $f$, from a source at a distance $D$, can be written as \citep{Owen:2010ng}:
\begin{equation}
\alpha = 0.028 \left( {h_0\over{10^{-24}}}\right )\left ( {D\over{1~\textrm{kpc}}}\right ) \left ({{\textrm{100~Hz}}\over f} \right )^3 
\label{eq:rmodes}
\end{equation}
Using this relation we convert the amplitude upper limits in upper limits on the r-mode amplitude, as shown in Fig. \ref{fig:alphaULs}.

\section{Discussion and Conclusions}
\label{sec:conclusions}

The G347.3 results presented here are the most sensitive to date across the frequency range:  Below 400 Hz this search improves by about 12\% on our previous results \citep{Ming_2022}, due to having followed-up 1 million further candidates. 
Above 400 Hz the improvement with respect to previous results is larger -- about 27\% with respect to our O1-data deep follow-up \citep{ Papa_2020midth}. Across the band our result is significantly more sensitive than that of \cite{LIGOScientific:2021mwx}, but in fairness we note that G347.3 is just one of 15 supernova remnants targeted there and the particular approach used for G347.3 based on a hidden Markov model \citep{Sun:2017zge} is very robust to possible deviations of the signal from the IT-{\it{2}} model, making these two searches complementary. 

The Vela Jr. results are between 12\%-17\% less constraining than those that \cite{ligo_o3a_c_v} obtained with O3 data. On the other hand the O3 data below 400 Hz is a factor of 1.6 more sensitive than the O2 data that we use here. We were able to compensate for this gap in raw-data sensitivity with a more sensitive search.

Since $\ddot{f}\in [0,7 {f/\tau^2}]$, rather than $[2|\fdot|^2/f, 7|\fdot|^2/f]$ as in \citep{ligo_o3a_c_v}, we survey a broader volume in frequency evolution space compared to the latest LVK searches. Our $\ddot{f}$ range is not just broader, but it also is independent of the value of the template's $\fdot$, so 
it accommodates signals that do not strictly follow a power law.

These null results can be used to constrain the amplitude of continuous gravitational waves from G347.3 and Vela Jr. An indirect, age-based, upper limit on the gravitational wave amplitude can also be computed and is useful to gauge whether a search has attained an interesting sensitivity level : 
If we assume that the neutron star has spun down all its life $\tau$ according to a power law with braking index $n$, $\dot{f} \propto f^n$, and if we assume that all the rotational energy loss is solely due to gravitational wave emission, the resulting gravitational wave amplitude at a distance $D$ is \citep{LIGOScientific:2008hqb}:

\begin{equation}
h_0^{\rm age} \le \frac{1}{D} \sqrt{\frac{n G I}{8 c^3 \tau}} \;,
\end{equation}
where $c$ is the speed of light, $G$ is the gravitational constant,  and $I=10^{38} \textrm{kg m}^2$ is the canonical moment of inertia of neutron star. Our searches are well below ($\approx$ by a factor 10) this limit  for all targets in most of the surveyed frequency range, for both $n=7$ (r-modes) and $n=5$ (equatorial deformations)  as shown in Fig.s~\ref{fig:ULs_V1} and \ref{fig:ULs_G1G2}.

Recast in terms of equatorial ellipticity, for G347.3 our results constrain it below $10^{-6}$ at frequencies higher than $\approx$ 290 Hz, dropping to $< 6\times 10^{-7}$ at 400 Hz and to $1.4\times 10^{-7}$ at 1300 Hz. Assuming a distance of 200 pc for Vela Jr., we constrain its ellipticity to be $< 10^{-7}$ at 400 Hz.
Values of the ellipticity that are a few $\times10^{-7}$ are completely plausible \citep{McDanielJohnsonOwen,Gittins:2020cvx,Gittins:2021zpv}.

 Targeted searches for emission from known pulsars have excluded much smaller ellipticities \citep{LVC_2019_targeted,LVC_2020_targeted,Ashok:2021fnj, LVC_2022_targeted}, so one could argue that we already know that neutron stars are extremely axis-symmetric. Pulsars are however generally much older than the young neutron stars targeted here. Vela Jr. and G347.3 are much closer to their birth than typical pulsars, and that birth process is a catastrophic event unlikely to be spherically symmetric. It is not unlikely that the newborn neutron star presents some leftover non-axisymmetry \citep{Janka_2008}. This is however expected to 
reduce in time as the star cools down  \citep{thm_decay,Ohmic_decay}, producing a population of very spherical older pulsars. 

Emission of continuous waves due to r-modes has long been thought to be a mechanism at work in fast-rotating newly-born neutron stars, that could explain why so many young neutron stars appear to spin much slower than the maximum rotation rates expected after the collapse \citep{Arras_2003}. General relativity predicts r-modes to grow under the emission of gravitational waves, i.e. to be unstable, with the growth hampered by viscosity, which depends on the neutron star structure and most importantly its temperature. The interplay between the rotation frequency, the r-modes amplitude and the temperature of the star is complex, but it is expected that a young neutron might be emitting through r-modes for a year - thousands of years after its birth \citep{Arras_2003,Owen:2010ng}. The amplitude is expected $\alpha \ll 1$ with estimates as low as $10^{-5}$ \citep{PhysRevD.76.064019} and $10^{-3}$ considered possibile, but large \citep{Haskell:2015iia}. Our r-mode amplitude constraints of  
$\alpha < 10^{-4}$ at frequencies higher than $\approx$ 290 Hz for G347.3 and above 150 Hz for Vela Jr. (assuming a distance of 200 pc) are hence probing physically meaningful values.

The detection of a continuous gravitational wave signal could happen with any search that breaks new territory. This has motivated the present investigation of 5 million sub-threshold candidates from very broad waveform-bank searches for emission from the young supernova remnants G347.3 and Vela Jr. The Stage 0 searches are the most computationally intensive and they were made possible thanks to the Einstein@Home volunteers, whom we express our heartfelt gratitude to. As the sensitivity of the data increases, the chances of identifying a signal also increase, and we eagerly wait for new data to be publicly released to continue to search.

\section{Acknowledgments}

We gratefully acknowledge the support of the many thousands of Einstein@Home volunteers who made this search possible. \\
We acknowledge support from the Max Planck Society for Projects QPQ10003 and QPQ10004, and the NSF grant 1816904.\\ 
A lot of post-processing is run on the ATLAS cluster at AEI Hannover. We thank Carsten Aulbert and Henning Fehrmann for their support. \\
We would like to thank the instrument-scientist and engineers of LIGO whose amazing work has produced detectors capable of probing gravitational waves so incredibly small.\\
This research has made use of data, software and/or web tools obtained from the Gravitational Wave Open Science Center (https://www.gw-openscience.org/ ), a service of LIGO Laboratory, the LIGO Scientific Collaboration and the Virgo Collaboration. LIGO Laboratory and Advanced LIGO are funded by the United States National Science Foundation (NSF) as well as the Science and Technology Facilities Council (STFC) of the United Kingdom, the Max-Planck-Society (MPS), and the State of Niedersachsen/Germany for support of the construction of Advanced LIGO and construction and operation of the GEO600 detector. Additional support for Advanced LIGO was provided by the Australian Research Council. Virgo is funded, through the European Gravitational Observatory (EGO), by the French Centre National de Recherche Scientifique (CNRS), the Italian Istituto Nazionale di Fisica Nucleare (INFN) and the Dutch Nikhef, with contributions by institutions from Belgium, Germany, Greece, Hungary, Ireland, Japan, Monaco, Poland, Portugal, Spain.

\appendix
\section{Clustering}
\label{App:clustering}

We use the density clustering method of \cite{den_cluster}. Several parameters  -- specified in Table 1 of \citep{den_cluster} -- determine the operation of the algorithm. Each set of clustering parameters results in a different number of candidates and in a different uncertainty region around each candidate. These in turn affect both the computing cost of the next stage and the sensitivity of the search, since any candidate that is not selected by the clustering is forever lost.

We select the clustering parameters so that at fixed computing cost of the Stage 1 follow-up search, the sensitivity of the search is maximised for our target signal population. With an allocation of about $9\times10^{8}$, $7\times10^{8}$ and $7\times10^{9}$  CPU seconds for Vela Jr., low frequency G347.3 and high frequency G347.3 follow-up searches respectively, the resulting clustering parameters are listed in Table \ref{tab:cluster}.

\begin{table*}[ht]
\centering
\begin{tabular}{|c|c|c|c|}
\hline
\hline
& Vela Jr. (20 - 400 Hz) & G347.3 (20 - 400 Hz) & G347.3 (400 - 1300 Hz) \\
\hline
\hline
 Input threshold $\Gamma_\mathrm{L}$ in $\BSGLtLr$ & -2.50& -2.00 & -11.50 \\
 \hline
 Output threshold $\Gamma_\mathrm{S}$ in $\BSGLtLr$ & 2.61 & 1.95 & 2.70\\
  \hline
Occupancy threshold $N_\mathrm{occ}$& 1 & 2 &2 \\
 \hline
 Binning width in $f$ ($\times  \delta{f}$) &  12  & 7 & 3 \\
  \hline
 Binning width in $\fdot$ ($\times \frac{\delta{\dot{f}}}{\gamma_1}$) & 11 &  4& 4 \\
  \hline
Binning width in $\fddot$ ($\times \frac{\delta{\ddot{f}}}{\gamma_2}$) & 8 &  3 & 2 \\
  \hline
 Neighboring occupancy summing  & Yes & Yes & Yes \\
 \hline

\hline  

\end{tabular}
\caption{Optimal density clustering parameters we determined for these searches using Monte Carlos.}
\label{tab:cluster}
\end{table*}

\newpage

\bibliography{paperBibApJ}
\bibliographystyle{aasjournal}

\newpage

\end{document}

%% file: macros.tex
\def\Tobs{T_{\textrm{\mbox{\tiny{obs}}}}}
\def\Tcoh{T_{\textrm{\mbox{\tiny{coh}}}}}
\def\Tref{T_{\textrm{\mbox{\tiny{ref}}}}}

\def\EatH{Einstein@Home}

\def\sci#1#2{#1\times10^{#2}}

\newcommand{\avgSeg}[1]{\overline{#1}}			

\newcommand{\Freq}{f}
\newcommand{\fdot}{{\dot{\Freq}}}
\newcommand{\fddot}{\ddot{\Freq}}

\newcommand{\M}{\mathcal{M}}

\newcommand{\Gauss}{\mathrm{\MakeUppercase{G}}}
\newcommand{\Signal}{{\mathrm{\MakeUppercase{S}}}}
\newcommand{\Line}{{\mathrm{\MakeUppercase{L}}}}
\newcommand{\Transient}{{\mathrm{t\MakeUppercase{L}}}}

\newcommand{\NoisetL}{{\Gauss\Line\Transient}}

\providecommand{\sc}[1]{\widehat{#1}}
\renewcommand{\sc}[1]{\widehat{#1}}

\newcommand{\BSNtsc}{{\hat\beta}_{{\Signal/\NoisetL}}}	
\newcommand{\BSGLtLr}{{\hat\beta}_{{\Signal/\NoisetL r}}}

\newcommand{\F}{\mathcal{F}}		

\newcommand{\avF}{\avgSeg{\F}}

\newcommand{\avMis}{\avgSeg{m}}

\newcommand{\Nseg}{{N_{\mathrm{seg}}}}

\newcommand{\posVelaJra}{\ensuremath{2.3213891342490}}
\newcommand{\posGa}{\ensuremath{4.509370536464}} 
\newcommand{\posVelaJrd}{\ensuremath{-0.8080542824176}} 
\newcommand{\posGd}{\ensuremath{-0.6951890756789}} 

\newcommand{\paramfdotloVela}{\ensuremath{-\sci{1.8}{-8}}} 
\newcommand{\paramfdothiVela}{\ensuremath{0}}
\newcommand{\paramfdothiG}{\ensuremath{0}} 
\newcommand{\paramfdotloG}{\ensuremath{-\sci{8.0}{-9}}} 
\newcommand{\paramfddotloVela}{\ensuremath{0}} 
\newcommand{\paramfddothiVela}{\ensuremath{\sci{5.7}{-18}}}
\newcommand{\paramfddothiG}{\ensuremath{\sci{1.1}{-18}}}
\newcommand{\paramfddotloG}{\ensuremath{0}} 

\newcommand{\TrefGPS}{\ensuremath{1177858472.0}}

\newcommand{\paramWUgputimemins}{20} 
\newcommand{\paramWUcputimeHours}{8} 
\newcommand{\paramtotalWUsmillionsGV}{3.1} 
\newcommand{\avgTempWU}{\ensuremath{\sci{3.1}{11}}}
\newcommand{\ThrVelalow}{2.61}

\newcommand{\ThrGlow}{1.95}
\newcommand{\ThrGmid}{2.70}

\newcommand{\NumVelalow}{\ensuremath{\sci{1.2}{6}}}

\newcommand{\NumGlow}{\ensuremath{\sci{1.0}{6}}}
\newcommand{\NumGmid}{\ensuremath{\sci{3.1}{6}}}

\newcommand{\RoneVlowFst}{2.0}
\newcommand{\RoneGlowFst}{1.8}

\newcommand{\RoneGmidFst}{2.0}

\newcommand{\NrejectFUoneVlowFst}{\ensuremath{\sci{6.4}{5}}}
\newcommand{\NrejectFUoneGlowFst}{\ensuremath{\sci{4.7}{5}}}

\newcommand{\NrejectFUoneGmidFst}{\ensuremath{\sci{1.4}{6}}}

\newcommand{\lowestULvOne}{$6.4\times 10^{-26}$}
\newcommand{\lowestULvOnefreq}{162.5}
\newcommand{\lowestULgOne}{$6.2\times 10^{-26}$}
\newcommand{\lowestULgOnefreq}{161.5}